\documentclass[fleqn,usenatbib]{mnras}

\usepackage{newtxtext,newtxmath}

\usepackage[T1]{fontenc}

\DeclareRobustCommand{\VAN}[3]{#2}
\let\VANthebibliography\thebibliography
\def\thebibliography{\DeclareRobustCommand{\VAN}[3]{##3}\VANthebibliography}

\usepackage{graphicx}	
\usepackage{amsmath}	
\usepackage{indentfirst}
\usepackage{siunitx}
\usepackage{CJKutf8}
\usepackage{color}
\usepackage[dvipsnames]{xcolor}
 \usepackage{subfig}

\numberwithin{equation}{section}
\usepackage{booktabs}
\usepackage{multirow}
\usepackage{multicol}

\title[Exploring IDE with CQPSO]{Exploring Interacting Dark Energy with Chaos Quantum-Behaved Particle Swarm Optimization}

\author[Z. Yin, Z. Ren, A. A. Costa]{
Zhixiang Yin,$^{1}$
Zelin Ren,$^{1}$
André A. Costa,$^{1,2}$\thanks{E-mail: andrecosta@yzu.edu.cn}
\\
$^{1}$Center for Gravitation and Cosmology, College of Physical Science and Technology, Yangzhou University, Yangzhou 225009, China \\
$^{2}$College of Physics, Nanjing University of Aeronautics and Astronautics, Nanjing 211106, China \\
}
\date{Accepted XXX. Received YYY; in original form ZZZ}

\pubyear{2023}

\begin{document}
\begin{CJK}{UTF8}{gkai}
\label{firstpage}
\pagerange{\pageref{firstpage}--\pageref{lastpage}}
\maketitle

\begin{abstract}
Models with an interaction between dark energy and dark matter have already been studied for about twenty years. However, in this paper, we provide for the first time a general analytical solution for models with an energy transfer given by $\mathcal{E} = 3H(\xi_1 \rho_c + \xi_2 \rho_d)$. We also use a new set of age-redshift data for 114 old astrophysical objects (OAO) and constrain some special cases of this general energy transfer. We use a method inspired on artificial intelligence, known as Chaos Quantum-behaved Particle Swarm Optimization (CQPSO), to explore the parameter space and search the best fit values. We test this method under a simulated scenario and also compare with previous MCMC results and find good agreement with the expected results.
\end{abstract}




\section{Introduction}
The standard cosmological model, the $\Lambda$CDM model, is the simplest one that can fit the current observational data. However, this model possesses some theoretical challenges known as {\it the cosmological constant problem} and {\it the coincidence problem}. The first one concerns the small observed value for the cosmological constant $\Lambda$, which is several orders of magnitude smaller than expected for a vacuum energy in a quantum field theory \citep{Weinberg:1988cp}. The second is related to the fact that although the energy densities of dark energy and dark matter evolve in a completely different way, they have the same order of magnitude nowadays \citep{Chimento:2003iea}. Motivated by the coincidence problem, it has been proposed an interaction in the dark sector (for a review, see \cite{Wang:2016lxa}).

In order to investigate the possibility of an interaction and constrain its parameters, several cosmological data have been used in the literature (e. g. \cite{Pellicer:2011mw,Costa:2013sva,Ferreira:2014jhn,Costa:2016tpb,Costa:2018aoy,An:2018vzw}, to cite a few). It has been realized that higher redshifts induce stronger deviations from the standard cosmology, if we fix the cosmological parameters to their values today \citep{Costa:2019uvk}. This means that a combination of low and high redshift data is crucial to impose strong constraints on interacting models.

The ages of old astrophysical objects (OAO) played an important role establishing the $\Lambda$CDM model, with reports of OAO being older than the Universe assuming the prevailing Einstein-de Sitter model \citep{Dunlop:1996mp,Jimenez:1996at,VandenBerg:1996tm}. This led to an age crisis \citep{Jaffe:1995qu,1995Natur376399B,Krauss:1995yb,Ostriker:1995su,Alcaniz:1999kr}, which was solved by the discovery of the late-time cosmic acceleration and the necessity of a dark energy component using SNIa \citep{SupernovaSearchTeam:1998fmf,SupernovaCosmologyProject:1998vns}. After the end of the age crisis, the OAO received less attention, but some authors still used them to constrain dark energy and other cosmological parameters \citep{Jimenez:2003iv,Capozziello:2004jy,Samushia:2009px,Dantas:2010zh,Verde:2013fva,Bengaly:2013afa,Wei:2015cva,Rana:2016gha,Nunes:2020yij,Borghi:2021rft}. In addition, some works have proposed using the age of OAO as an independent route to investigate the Hubble tension \citep{Jimenez:2019onw,Bernal:2021yli,Boylan-Kolchin:2021fvy,Krishnan:2021dyb,Vagnozzi:2021tjv,Wei:2022plg,Costa:2023cmu}.

Recently a new compilation of high-z OAO with their associated age-redshift relation became available \citep{Vagnozzi:2021tjv}. This new data can be used to infer their lookback time and constrain the background parameters in an interacting dark energy model. In this work, we use an artificial intelligence inspired method, known as {\it Chaos Quantum-behaved Particle Swarm Optimization} (CQPSO) \citep{Sun,Lin}, to explore the parameter-space and search the best fit values. This method is an extension of the more traditional PSO algorithm \citep{kennedy_1,kennedy_2,engelbrecht}, which was previously also used for cosmological estimations \citep{Prasad:2011rd}. However, PSO methods just give the location of the best fit point. Therefore, we combine the Fisher matrix formalism to infer the covariance matrix and constrain our cosmological parameters \citep{Tegmark:1996bz,Coe:2009xf}.

This paper is organized as follows. In Section~\ref{sec:IDE}, we present our phenomenological interacting dark energy models and obtain a general analytical solution for their energy densities. In Section~\ref{sec:LBT}, we discuss the use of the ages of old astrophysical objects to estimate their lookback time and show its likelihood. Then, we describe the particle swarm optimization method and some extensions to obtain the best fit in Sect.~\ref{sec:PSO}. The Fisher matrix formalism to infer the cosmological constraints is shown in Sect.~\ref{sec:Fisher}. We test our routines with some full controlled simulations in Sect.~\ref{sec:Simulation}, which can also be used as a forecast for future constraints. Finally, Sect.~\ref{sec:Real} presents our constraints from real data and a comparison with previous estimations. We summarize our results and present our conclusions in Sect.~\ref{sec:Conclusions}.

\section{Interacting dark energy}
\label{sec:IDE}
\subsection{Evolution equations}
In a homogeneous and isotropic universe, the components can be treated as a perfect fluid with energy-momentum tensor given by
\begin{equation}
\label{eq.2.1}
\mathcal{T}_{\mu \nu} = (\rho+p)u_{\mu}u_{\nu}+ g_{\mu \nu}p \,,
\end{equation}
where $\rho$ is the energy density, $p$ is the pressure, $u_{\mu}$ is the four-velocity vector, and $g_{\mu \nu}$ is the Friedmann-Lemaitre-Robertson-Walker (FLRW) metric. In a matrix form, this metric can be written as
\begin{equation}
g_{\mu \nu} = 
\begin{pmatrix}
-1 & 0  & 0  & 0 \\
 0 & \frac{a^2(t)}{1-Kr^2} & 0 & 0 \\
 0 & 0 & a^2(t)r^2 & 0 \\
 0 & 0 & 0 & a^2(t)r^2sin^2\theta
\end{pmatrix} \,,
\end{equation}
where $a(t)$  is a scale factor accounting for the expansion or contraction of the universe and $K$ is a constant which establishes the geometry of the spatial section.

Hence, from the Einstein's field equations
\begin{equation}
\mathcal{R}_{\mu \nu} - \frac{1}{2}g_{\mu \nu}\mathcal{R} = 8\pi G \mathcal{T}_{\mu \nu},
\end{equation}
we can obtain the Friedmann's equations
\begin{equation}
\label{eq.2.4}
\begin{aligned}
H^2(t) &=\frac{8\pi G}{3}\rho(t)-\frac{K}{a^2(t)} \,, \\
\Dot{H}(t) &= -4\pi G[\rho(t)+p(t)] + \frac{K}{a^2(t)} \,,
\end{aligned}
\end{equation}
where $H \equiv \frac{\Dot{a}(t)}{a(t)}$ denotes the Hubble parameter. We can also combine Eqs.~(\ref{eq.2.4}) and eliminate $K$ and $\Dot{H}$. The resultant equation describes the evolution of the energy density for all components
\begin{equation}
\label{eq.2.5}
\Dot{\rho}+3H(\rho+p) = 0 \,.
\end{equation}

In an interacting dark energy (IDE) model, we assume an energy transfer $\mathcal{E}$ between dark energy and dark matter. Therefore, the continuity equation for dark matter (c) and dark energy (d) are respectively given by
\begin{equation}
\label{eq.2.6}
\begin{aligned}
&\Dot{\rho}_{c} +3H\rho_{c} = + \mathcal{E}, \\
&\Dot{\rho}_{d} +3H(1+\omega)\rho_{d} = - \mathcal{E}
\end{aligned}
\end{equation}
where $\omega = \frac{p_{d}}{\rho_{d}}$ is the equation of state of dark energy. We suppose the energy transfer $\mathcal{E}$ is proportional to a combination of the energy densities of the dark sectors. Following \cite{He:2008si}, the energy transfer at linear order can be written in general as $\mathcal{E} = 3H(\xi_1\rho_c + \xi_2\rho_d)$, where $\xi_1$ and $\xi_2$ are constants and parameterize the interaction between the dark sectors.

\subsection{Solutions via matrix exponential functions}
The evolution equations for dark energy and dark matter form a first-order linear homogeneous system of differential equations. It is quite not easy to solve this system of equations even though it seems simple. Therefore, the solution has been obtained previously in the literature only for some special cases, such as $\xi_1 = 0$ or $\xi_2 = 0$ \citep{Bachega:2019fki}. And more general solutions have been obtained only numerically. In this section, we will briefly discuss about a simple method to solve the equations and present the general solution for these IDE models.

\subsubsection{Method using matrix exponential functions}
Consider solving a first-order linear homogeneous system of differential equations
\begin{equation}
\label{eq.2.7}
\left\{
\begin{aligned}
\Dot{x}_1 &= a_{11}x_1+a_{12}x_2+\dots+a_{1n}x_n, \\
\Dot{x}_2 &= a_{21}x_1+a_{22}x_2+\dots+a_{2n}x_n, \\
\vdots\\
\Dot{x}_m &= a_{m1}x_1+a_{m2}x_2+\dots+a_{mn}x_n. \\
\end{aligned}
\right.
\end{equation}
Initially, we let $\Vec{x} = (x_1, x_2, \dots, x_n)^T$ and
$\mathbf{A}=
\begin{pmatrix}
a_{11}& \dots & a_{1n} \\
\vdots& \ddots & \vdots \\
a_{m1}& \dots & a_{mn}
\end{pmatrix}$. Therefore, we can simplify Eq.~(\ref{eq.2.7}) as
$\Dot{\Vec{x}}=\mathbf{A} \Vec{x}$.
Hence we can obtain the solution 
\begin{equation}
\label{eq.solution}
\Vec{x}=e^{\mathbf{A}t}\Vec{x_0}\,,
\end{equation}
where $e^{\mathbf{A}t}$ is called the matrix exponential function which can be expanded into a Maclaurin series as $e^{\mathbf{A}t}=\sum\limits_{n=0}\limits^{\infty}\frac{t^n}{n!}\mathbf{A}^n$ . 

Furthermore, since the Laplace transform
\begin{equation}
\begin{aligned}
\mathscr{L}(e^{\mathbf{A}t}) &= \int^\infty_0 e^{\mathbf{A}t} \cdot e^{-st\mathbf{E}}dt = \int^\infty_0 e^{-(s\mathbf{E}-\mathbf{A})t} dt \\
&= (s\mathbf{E}-\mathbf{A})^{-1} \,,   
\end{aligned}
\end{equation}
it follows that $e^{\mathbf{A}t} = \mathscr{L}^{-1}[(s\mathbf{E}-\mathbf{A})^{-1}],$ where $\mathscr{L}^{-1}$ is the Laplace inverse transform and $\mathbf{E}$ is a unit matrix. Finally, the solution of Eq.~(\ref{eq.2.7}), i.e. Eq.~(\ref{eq.solution}), is obtained as follows
\begin{equation}
\Vec{x}={\mathscr{L}}^{-1}[(s\mathbf{E}-\mathbf{A})^{-1}] \Vec{x_0} \,,
\end{equation}
where $\Vec{x_0}$ denotes the initial value vector for the independent variables.

\subsubsection{The solution of the evolution equations for IDE models}
We first consider the most general case where the energy transfer between the dark sectors is proportional to the linear combination of their energy density. In this case, the evolution equations become
\begin{equation}
\label{eq.2.10}
\left\{
\begin{aligned}
&\Dot{\rho}_{c} +3H\rho_{c} = 3H(\xi_1\rho_c + \xi_2\rho_d) \,, \\
&\Dot{\rho}_{d} +3H(1+\omega)\rho_{d} = -3H(\xi_1\rho_c + \xi_2\rho_d) \,.
\end{aligned}
\right.
\end{equation}
Due to the factor of $H=\frac{\Dot{a}}{a}$, we first make the coordinate transformation $\tau = \ln(a)$ to write Eq.~(\ref{eq.2.10}) into the form of Eq.~(\ref{eq.2.7}) as
\begin{equation}
\label{eq.2.11}
\left\{
\begin{aligned}
&\frac{d\rho_{c}}{d\tau} = 3(\xi_1-1)\rho_{c} + 3\xi_2\rho_d \,, \\
&\frac{d\rho_{d}}{d\tau} =-3\xi_1\rho_c -3(1+\omega+\xi_2)\rho_{d} \,,
\end{aligned}
\right.
\end{equation}
where the coefficient matrix is
$\mathbf{A} = 
\begin{pmatrix}
    3(\xi_1-1) & 3\xi_2 \\
    -3\xi_1 & -3(1+\omega+\xi_2)
\end{pmatrix}$.

Hence,  
\begin{equation}
\label{eq.sE-A}
(s\mathbf{E}-\mathbf{A})^{-1} =
\begin{pmatrix}
    \frac{s+3(1+\omega+\xi_2)}{(s-s_1)(s-s_2)} & \frac{3\xi_2}{(s-s_1)(s-s_2)} \\
    \frac{-3\xi_1}{(s-s_1)(s-s_2)} & \frac{s-3(\xi_1-1)}{(s-s_1)(s-s_2)}
\end{pmatrix} \,,
\end{equation}
where $s_1=-\frac{3}{2}(2+\omega+\xi_2-\xi_1-\sqrt{\Delta})$ and $s_2=-\frac{3}{2}(2+\omega+\xi_2-\xi_1+\sqrt{\Delta})$ are the eigenvalues of the coefficient matrix $\mathbf{A}$, and $\Delta=(\omega+\xi_1+\xi_2)^2-4\xi_1\xi_2$. We suppose $\Delta >0$, in which case each pole of the components in Eq.(\ref{eq.sE-A}) is a first-order pole. Accordingly we can easily calculate their Laplace inverse transform via residue theorem and obtain
\begin{align}
\begin{pmatrix}
    \rho_c \\
    \rho_d
\end{pmatrix}
=
\begin{pmatrix}
    \frac{1}{2}[(1+\mathscr{A})a^{s_1}+(1-\mathscr{A})a^{s_2}] \quad \frac{\xi_2}{\sqrt{\Delta}}(a^{s_1}-a^{s_2}) \\
    -\frac{\xi_1}{\sqrt{\Delta}}(a^{s_1}-a^{s_2}) \quad \frac{1}{2}[(1-\mathscr{A})a^{s_1}+(1+\mathscr{A})a^{s_2}]
\end{pmatrix}
\begin{pmatrix}
    \rho_{c0} \\
    \rho_{d0}
\end{pmatrix} \,,
\end{align}
where $\mathscr{A}=\frac{\omega+\xi_1+\xi_2}{\sqrt{\Delta}}$. Considering the redshift $z$ and the scale factor $a$ obey the relation $(1+z) = a_0/a$,
and normalizing such that $a_0 = 1$, we obtain the final solution for Eq.~(\ref{eq.2.11}) as
\begin{equation}
\begin{aligned}
\label{eq.general}
\rho_c &= \frac{1}{2}[(1+\mathscr{A})(1+z)^{-s_1}+(1-\mathscr{A})(1+z)^{-s_2}]\rho_{c0}\\
&+\frac{\xi_2}{\sqrt{\Delta}}[(1+z)^{-s_1}-(1+z)^{-s_2}]\rho_{d0} \,, \\
\rho_d &= -\frac{\xi_1}{\sqrt{\Delta}}[(1+z)^{-s_1}-(1+z)^{-s_2}]\rho_{c0}\\
&+\frac{1}{2}[(1-\mathscr{A})(1+z)^{-s_1}+(1+\mathscr{A})(1+z)^{-s_2}]\rho_{d0} \,.
\end{aligned}
\end{equation}

Following previous works in the literature, we consider three special cases of this general solution as shown in Table~\ref{tab.models} \citep{Costa:2013sva,Costa:2016tpb,Bachega:2019fki}. Model I corresponds to the special case with $\xi_1 = 0$. In this case, $\Delta=(\omega+\xi_2)^2$, $s_1=-3$, $s_2=-3(1+\omega+\xi_2)$, and $\mathscr{A}=1$. Therefore, Eq.~(\ref{eq.general}) simplifies to
\begin{equation}
\label{eq:model_I}
\begin{aligned}
\rho_c &= (1+z)^3\rho_{c0} + \frac{\xi_2}{\omega+\xi_2}[(1+z)^3-(1+z)^{3(1+\omega+\xi_2)}]\rho_{d0} \,,\\
\rho_d &= (1+z)^{3(1+\omega+\xi_2)}\rho_{d0} \,. \\
\end{aligned}
\end{equation}
And the condition $\Delta=(\omega+\xi_2)^2 > 0$ implies $\omega+\xi_2 \neq 0$.
\begin{table}
    \centering
    \caption{Interacting dark energy models.}
    \begin{tabular}{|c|c|}
    \hline
        Model & $\mathcal{E}$ \\ \hline
        I & 3 $\xi_2$ $H$ $\rho_d$ \\ \hline
        II & 3 $\xi_1$ $H$ $\rho_c$ \\ \hline
        III & 3 $\xi$ $H$ ($\rho_c$+$\rho_d$) \\ \hline
    \end{tabular}
    \label{tab.models}
\end{table}

Similarly, we can derive the solution for Model II setting $\xi_2=0$. The interim parameters become $\Delta=(\omega+\xi_1)^2$, $s_1=-3(1-\xi_1)$, $s_2=-3(1+\omega)$, and $\mathscr{A}=1$. The equations for the energy densities of dark matter and dark energy are then obtained as follows
\begin{equation}
\label{eq:model_II}
\begin{aligned}
\rho_c &= (1+z)^{3(1-\xi_1)}\rho_{c0} \,, \\
\rho_d &= -\frac{\xi_1}{\omega+\xi_1}[(1+z)^{3(1+\xi_1)} - (1+z)^{3(1+\omega)}] \rho_{c0} \\
 &+ (1+z)^{3(1-\xi_1)}\rho_{d0} \,,
\end{aligned}
\end{equation}
where $\omega + \xi_1 \neq 0$. Comparing Eqs.~(\ref{eq:model_I}) and (\ref{eq:model_II}) with the solutions in \cite{Bachega:2019fki}, we find the linear homogeneous system of differential equations can be solved correctly via matrix exponential functions.

Likewise, if we set $\xi_1=\xi_2=\xi$, we can derive the solution for Model III, which \cite{Bachega:2019fki} did not provide, as follows
\begin{equation}
\begin{aligned}
&\rho_c =\frac{\xi}{\sqrt{\Delta}}[(1+z)^{\frac{3(2+\omega-\sqrt{\Delta})}{2}}-(1+z)^{\frac{3(2+\omega+\sqrt{\Delta})}{2}}]\rho_{d0} \\
&+\frac{1}{2}[(1-\frac{\omega+2\xi}{\sqrt{\Delta}})(1+z)^{\frac{3(2+\omega+\sqrt{\Delta})}{2}}+(1+\frac{\omega+2\xi}{\sqrt{\Delta}})(1+z)^{\frac{3(2+\omega-\sqrt{\Delta})}{2}}]\rho_{c0} \,, \\
&\rho_d = -\frac{\xi}{\sqrt{\Delta}}[(1+z)^{\frac{3(2+\omega-\sqrt{\Delta})}{2}}-(1+z)^{\frac{3(2+\omega+\sqrt{\Delta})}{2}}]\rho_{c0}\\
&+\frac{1}{2}[(1+\frac{\omega+2\xi}{\sqrt{\Delta}})(1+z)^{\frac{3(2+\omega+\sqrt{\Delta})}{2}}+(1-\frac{\omega+2\xi}{\sqrt{\Delta}})(1+z)^{\frac{3(2+\omega-\sqrt{\Delta})}{2}}]\rho_{d0} \,,
\end{aligned}
\end{equation}
where $\Delta = \omega(\omega+4\xi)>0$. 

Therefore, these analytical expressions greatly simplify the calculation for the evolution of the energy densities of interacting dark energy and dark matter, which had only been obtained numerically for the most general scenarios. This also improves the computational time, as those expressions only depend on the initial values and some known cosmological parameters, and numerically we need loops and interpolations to properly match the functions in the coupled system. The analytical expression also makes it clear the physical behaviour of the system.

\section{Lookback time}
\label{sec:LBT}

\begin{figure}
\includegraphics[width=\columnwidth]{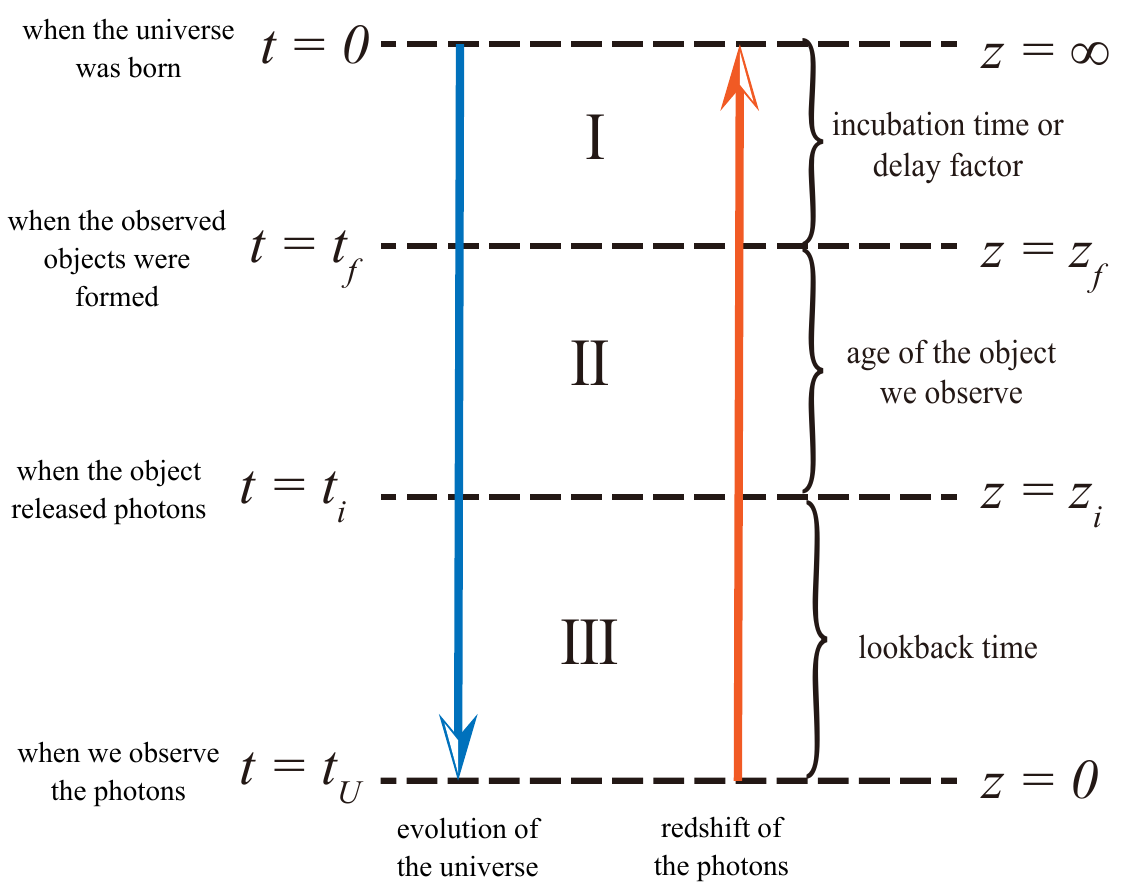}
\caption{Schematic view for the evolution of the universe and the redshift at several stages of old astrophysical objects.}
\label{fig.1}
\end{figure}

In order to constrain the parameters of these interacting dark energy models, we will use recent data for the ages of old astrophysical objects and their corresponding redshifts \citep{Vagnozzi:2021tjv}. The total time $\mathscr{t}$ spent by a light ray travelling from time $t_1$ to time $t_2$ can be calculated as 
%
%
%
\begin{equation}
\label{eq.3.1}
\mathscr{t}=\int^{t_2}_{t_1}dt=\int^{a_2}_{a_1}\frac{da}{aH(a)} \,,
\end{equation}
where we use the definition of the Hubble parameter $H=\Dot{a}/a$. Differentiating the redshift-scale factor relation, we obtain $dz = -\frac{da}{a^2}$. Therefore, combining this result with Eq.~(\ref{eq.3.1}), we can derive the relationship between the total time and redshift as
\begin{equation}
\label{eq.3.2}
\mathscr{t} = \int^{z_1}_{z_2}\frac{dz}{(1+z)H(z)} \,,
\end{equation}
where $z_1$ and $z_2$ are the redshifts at time $t_1$ and $t_2$, respectively.

Figure~\ref{fig.1} shows us that the redshift $z_1=\infty$ at the beginning of the universe. Hence, we can calculate the age of the universe at redshift $z_2=z_i$ as
\begin{equation}
\label{eq.3.3}
\mathscr{t}_U(z_i) = \int^{\infty}_{z_i}\frac{dz}{(1+z)H(z)}.
\end{equation}
On the other hand, the lookback time at redshift $z_i$ can be theoretically obtained by
\begin{equation}
\label{eq.3.4}
\begin{aligned}
\mathscr{t}_L(z_i) =\int^{z_i}_{0}{\frac{dz}{(1+z)H(z)}} \,,
\end{aligned}
\end{equation}
which denotes the difference between the current age of the universe $\mathscr{t}_U$ and the age $\mathscr{t}_i$ at redshift $z_i$. Therefore, the age of an object we observed at redshift $z_i$ is
\begin{equation}
\label{eq.3.5}
\begin{aligned}
\mathscr{t}_{age}(z_i) =\int^{z_f}_{z_i}{\frac{dz}{(1+z)H(z)}} = \mathscr{t}_L(z_F) - \mathscr{t}_L(z_i) \,,
\end{aligned}
\end{equation}
where $z_F$ is the redshift corresponding to the moment of formation of the object. Finally, combining these equations, the observed lookback time $\mathscr{t}^{obs}_L(z_i)$ to the object at redshift $z_i$ is
\begin{equation}
\label{Eq.3.5}
\begin{aligned}
\mathscr{t}^{obs}_L(z_i)  &= \mathscr{t}_L(z_F) - \mathscr{t}(z_i) \\
&= [\mathscr{t}^{obs}_U - \mathscr{t}_{age}(z_i)] - [\mathscr{t}^{obs}_U - \mathscr{t}_L(z_F)] \\
 &= \mathscr{t}^{obs}_U - \mathscr{t}_{age}(z_i) - df \,,
\end{aligned}
\end{equation}
where $\mathscr{t}^{obs}_U$ is the observed age of the Universe and $df$ is the delay factor defined by $df\equiv \mathscr{t}^{obs}_U - \mathscr{t}_L(z_F)$.

The comparison between our theoretical estimation for the lookback time and the inferred values from the ages of old astrophysical objects can be realized assuming a Gaussian likelihood $\mathcal{L}$ given by
\begin{equation}
\label{Eq.3.6}
\begin{aligned}
\chi^2 &= -2 \ln(\mathcal{L}) \\
&= \frac{[\mathscr{t}_U(\Vec{\Theta}) - \mathscr{t}^{obs}_U]^2}{\sigma_{_U}^2} + \sum_{i=1}^n \frac{[\mathscr{t}_L(z_i,\Vec{\Theta})-\mathscr{t}_L^{obs}(z_i,df)]^2}{\sigma _i^2 + \sigma_{_U}^2} \,,
\end{aligned}
\end{equation}
where $\Vec{\Theta}$ represents our set of cosmological parameters.

\section{Particle swarm optimization}
\label{sec:PSO}
Our objective is determining the region in the parameter space which is in better agreement with the cosmological data. This can be obtained by sampling the parameter space and calculating the likelihood function. Points with larger likelihood are more probable and in better agreement with the data. Although straightforward, the brute-force method requires calculating the likelihood function for thousands of thousands of points in a D-dimensional parameter space. Therefore, it is usually used other sampling methods, such as Markov chain Monte Carlo (MCMC), which do not require calculating the likelihood in the whole space, but still holds information about its behaviour.

MCMC methods have successfully been applied to cosmological parameter estimation problems in the literature. However, an easier and faster method, both in terms of implementation and running time, can be obtained if we just look for the best fit value. Essentially, our task is to identify the parameter values that maximize the likelihood or, similarly, minimize the $\chi^2$ value. This exercise fundamentally falls within the realm of an optimization problem.


The term ''optimization'' refers to the process of identifying the best or most favorable solution within a set of feasible options for a given problem. The typical structure of an optimization problem is expressed as:
\begin{equation}
\begin{aligned}
\label{eq.optimal}
Min\{f(\Vec{X})\,, \forall \, \Vec{X} \in \mathcal{S} \} \,,
\end{aligned}
\end{equation}
where $\mathcal{S}$ is a $D$-dimensional parameter space. For the maximization problem, we can consider $g(x)=-f(x)$ converting it into the problem depicted by Eq.~(\ref{eq.optimal}). There exist numerous methods to tackle such problems. Among these, the particle swarm optimization (PSO) algorithm and its enhanced variants hold a prominent position.

\subsection{Particle swarm optimization algorithm}

The PSO algorithm was initially introduced by James Kennedy and Russell Eberhart in 1995 \citep{kennedy_1}, drawing inspiration from the collective behavior of birds. To introduce this algorithm and its variations, let us begin with a scenario where a group of birds ventures out in the morning to search for food. Upon returning to their nest, they engage in discussions about the locations with abundant food sources, which subsequently impacts the trajectory each bird takes in its quest for food the following day, until they finally find where exists the most abundant food. We then posit that the route of each bird can be described by the following formula:
$$
\left\{
\begin{aligned}
V_{i,j}(t+1)&= wV_{i,j}(t)\\
             &+c_1r_1(P_{i,j}(t)-X_{i,j}(t))+c_2r_2(G_{j}(t)-X_{i,j}(t))\,,\\
X_{i,j}(t+1) &= X_{i,j}(t) + V_{i,j}(t+1) \,.
\end{aligned}
\right.
$$

In the given equation, $X_{i,j}$ and $V_{i,j}$ represent the position and speed of the $i$-th bird in the $j$-th direction and $t$ is the count for today, consequently $t+1$ stands for the following day. The parameter $w$ is referred to as the "inertia factor" which signifies the level of trust the birds place in their current search patterns. Meanwhile, $c_1$ and $c_2$ are factors that gauge the degree of reliance each bird has on its own experience and the experiences of others, respectively. The terms $P_{i,j}(t)$ and $G_{j}(t)$ denote the positions where each bird has found the most food and where the collective group of birds have located the most food up to the current time. Finally, $r_1$ and $r_2$ serve as random parameters that express the subjective initiative of each individual bird.

The PSO algorithm is notable for its simplicity in calculations and ease of implementation. However, the standard PSO algorithm has been shown to struggle with converging to global optimal values \citep{10.5555/935867}. As a result, enhancing the particle swarm optimization approach has emerged as a prominent research focus.

Among the researchers in this field, Sun Jun has made significant contributions. He systematically introduced a quantum-behaved particle swarm optimization (QPSO) algorithm based on the $\delta$ potential well \citep{Sun}, inspired by Prof. Maurice Clerc's paper \citep{785513}. 
Subsequently, their team incorporated chaos theory into the QPSO algorithm, giving rise to the chaos quantum-behaved particle swarm optimization (CQPSO) algorithm \citep{Lin}. In this section, we will provide a brief overview of these two algorithms.

\subsection{Quantum-behaved particle swarm optimization algorithm}

\subsubsection{Establishment of the $\delta$ potential well model}
For simplicity, we consider a one-dimensional scenario with a potential well described by $U(X) = -\xi \delta(X-p),$ where $p$ signifies the position of the potential well. In this context, the corresponding Hamiltonian operator is $\hat{H}=-\frac{\hbar^2}{2m}\frac{d^2}{dX^2}-\xi \delta(X-p).$ Consequently, the time-independent Schrödinger equation takes the form:
\begin{equation}
\label{eq.4.2}
\frac{d^2\psi}{dX^2}+\frac{2m}{\hbar^2}[E+\xi\delta(X-p)]\psi=0 \,.
\end{equation}

The normalized wave function solution of Eq.~(\ref{eq.4.2}) is straightforward, as demonstrated in \cite{Sun}, and is given by
\begin{equation}
\label{eq.4.3}
\psi(y)=\frac{1}{\sqrt{L}}e^{-\frac{|y|}{L}} \,,
\end{equation}
where $y = X - p$ and $L=\frac{\hbar^2}{m\xi}$ stands for the characteristic length of the potential well. Consequently, the probability density function can be derived as
\begin{equation}
\label{eq.4.4}
pdf(y) = |\psi(y)|^2=\frac{1}{L}e^{-2|y|/L} \,,
\end{equation}
leading to the probability distribution function
\begin{equation}
\label{eq.4.5}
u(y) = e^{-2|y|/L} \,.
\end{equation}

\subsubsection{The evolution function of particles}
Reorganizing Eq.~(\ref{eq.4.5}), we can solve for the particle's position as follows:
\begin{equation}
\label{eq.4.6}
X(t) = p + y = p(t) \pm \frac{L}{2}\ln\left(\frac{1}{u(t)}\right) \,,
\end{equation}
where $u$ is a random number uniformly distributed within the interval (0, 1), i.e., $u \sim U(0,1)$. Extending Eq.~(\ref{eq.4.6}) to the scenario of $N$ particles in $D$ dimensions, we have:
\begin{equation}
\label{eq.4.7}
X_{i,j}(t) = p_{i,j}(t) \pm \frac{L_{i,j}}{2}\ln\left(\frac{1}{u_{i,j}(t)}\right) \,,
\end{equation}
where $i = 1,2,\dots, N$ and $j = 1,2,\dots, D$. This serves as the fundamental evolution function of the quantum-behaved particle swarm optimization for the $i$-th particle in the $j$-th dimension. If we let $P_{i,j}$ represent the local best position of the $i$-th particle in the $j$-th dimension, and $G_j$ denote the global best position, then the attractor $p_{i,j}$ in Eq.~(\ref{eq.4.7}) can be calculated as:
\begin{equation}
p_{i,j}(t) = \phi_j(t)P_{i,j}(t) + [1-\phi_j(t)]G_j(t),
\end{equation}
where $\phi_j(t)\sim U(0,1)$.

S. Jun emphasized that to ensure the convergence of the QPSO algorithm, it is advantageous to integrate the mean best position, denoted as $\Vec{M}(t)$, into the algorithm as follows:
\begin{equation}
\begin{aligned}
\label{eq.4.9}
\vec{M}(t)&=(M_1(t),M_2(t),\dots, M_D(t))\\
&=(\frac{1}{N}\sum^N_{i=1}P_{i,1}(t),\frac{1}{N}\sum^N_{i=1}P_{i,2}(t),\dots, \frac{1}{N}\sum^N_{i=1}P_{i,D}(t)) \,,
\end{aligned}
\end{equation}
and $L_{i,j}(t)$ can be calculated as
\begin{equation}
L_{i,j}(t) = 2\gamma|M_j(t)-X_{i,j}(t)| \,,
\end{equation}
where $\gamma$ is referred to as the contraction-expansion factor. This is the only parameter in the QPSO apart from the number of particles $N$ and the number of iterations.

In summary, the final evolution function of the quantum-behaved particle swarm optimization algorithm is described by the following set of equations:
\begin{equation}
\left\{
\begin{aligned}
\label{eq.4.11}
&p_{i,j}(t) = \phi_j(t)P_{i,j}(t)+[1-\phi_j(t)]G_j(t) \,,\\
& X_{i,j}(t+1) = p_{i,j}(t) \pm \gamma|M_j(t)-X_{i,j}(t)|\ln[1/u_{i,j}(t)] \,,
\end{aligned}
\right.
\end{equation}
where $\phi_j(t)\sim U(0,1)$ and $u_{i,j}(t)\sim U(0,1)$.

\subsection{Chaos quantum-behaved particle swarm optimization algorithm}
\label{sec:chaos}
Due to the sensitivity of the evolution of chaotic systems to initial conditions, Sun's research team incorporated chaos into the QPSO algorithm \citep{Lin}. This addition aimed to enhance the randomness of position transformations and decrease the probability that the algorithm becomes trapped in local maximum. To achieve this, they utilized the Logistic equation,
\begin{equation}
\label{eq.logistic}
\mu_{i+1}=4\mu_i(1-\mu_i)
\end{equation}
as the chaotic equation, where $0 \leq \mu_0 \leq 1$.

In order to streamline and enhance the code, we modified certain sections of the algorithm as compared with \cite{Lin}, but still following the same reasoning. 
Initially, with the growing number of iterations, the particles' positions tend to become increasingly close. This characteristic can be exploited to generate a chaotic sequence. In order to accomplish that, we first define a new position coordinate given by
\begin{equation}
\label{eq.4.13}
X^\mu_{i,j} = \frac{X_{i,j}-X^{min}_j}{X^{max}_j-X^{min}_j},
\end{equation}
where $X^{min}_j$ and $X^{max}_j$ denote the minimum and maximum values in the $j$-th dimension of the parameter space, respectively, while $X_{i,j}$ represents the $j$-dimensional coordinate of the $i$-th particle obtained from Eq.~(\ref{eq.4.11}).

Secondly, we insert the new coordinates $\{X^\mu_{i,j}\}$ into the chaotic equation Eq.~(\ref{eq.logistic}) and iterate several times to derive a new chaotic sequence $\{X^{\prime \mu}_{i,j}\}$.

Thirdly, we solve for the positions of particles $\vec{X}^{chaos}_{i}$ after undergoing chaotic evolution using the following formula:
\begin{equation}
\label{eq.4.14}
X^{chaos}_{i,j}= X^{min}_j +(X^{max}_j-X^{min}_j) \times X^{\prime \mu}_{i,j}.
\end{equation}

Fourthly, we compare the fitness of the chaotic particles $f(\vec{X}^{chaos}_{i})$ with the corresponding local best value $Pbest_i$, which is defined by $Pbest_i = f(\vec{P}_i)$ and so does $Gbest$ in the following context. If the condition $f(\vec{X}^{chaos}_{i}) < Pbest_i$ is satisfied, we update the position and fitness of the latter using the values of the former. However, in cases where the condition is not met, the effectiveness of chaotic optimization might be compromised. Therefore, to take full advantage of the points generated by chaos, we draw inspiration from the simulated annealing algorithm. When the condition is not satisfied, we adopt a probabilistic approach to updating. Specifically, the value of the former is assigned to the latter with a certain probability, given by:
\begin{equation}
\label{eq.4.15}
\mathcal{P} = e^{-\frac{\Delta f}{T}},
\end{equation}
where $\Delta f = f(\vec{X}^{chaos}_{i}) - Pbest_i$, which in this context will be positive. Additionally, $T$ is a temperature-like parameter defined by $T(t+1)=T(t) \times \alpha$ and represents the temperature that decreases exponentially with each iteration. We usually can set $T(0)=1000$ and $\alpha=0.99$.

\subsection{Code's workflow}
\begin{figure*}
\includegraphics[width=0.8\textwidth]{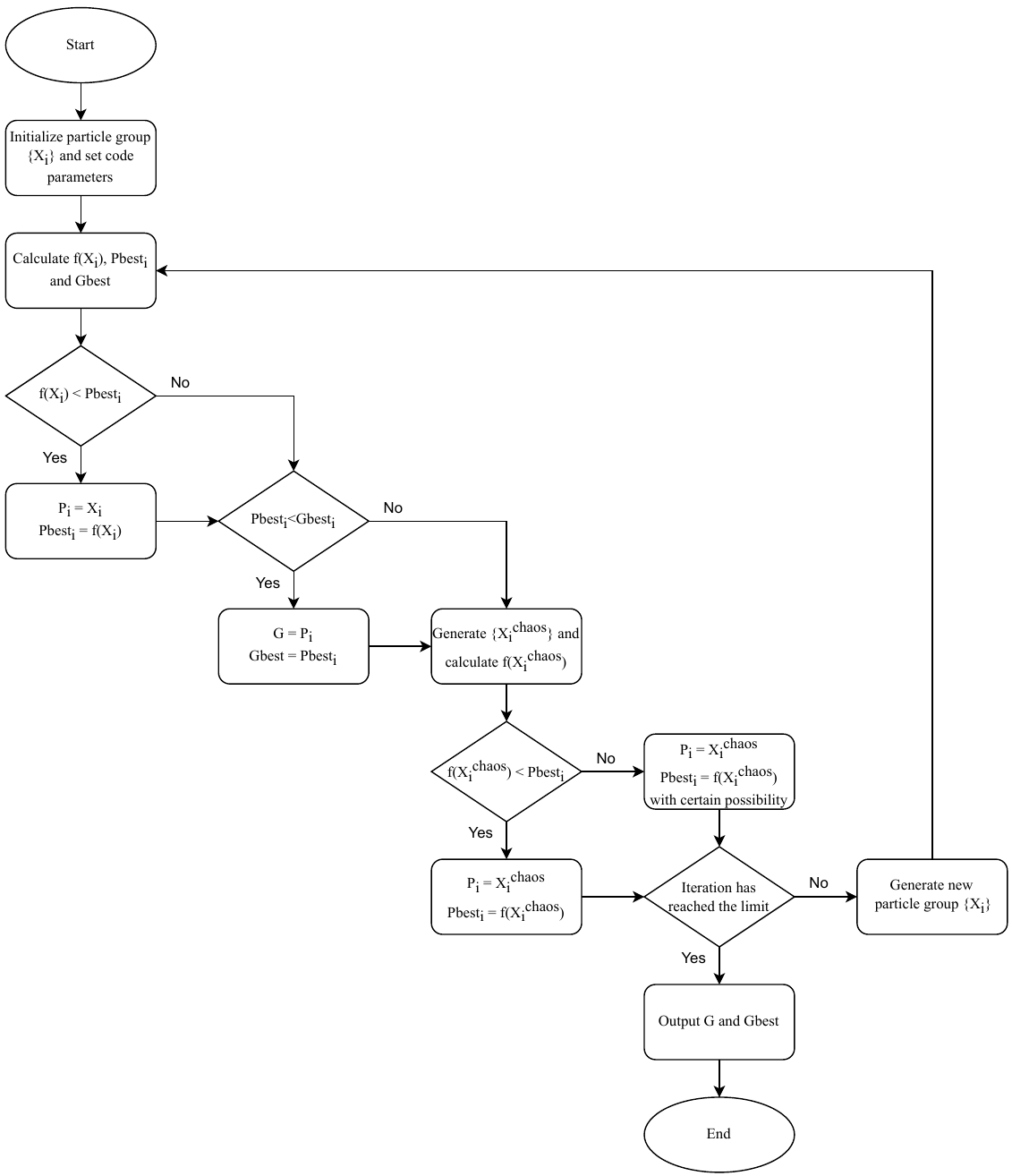}
\caption{Flowchart for the CQPSO algorithm.}
\label{fig.flow chart}
\end{figure*}

In this section, we provide a concise overview of the code's workflow, which is encapsulated in the flowchart depicted in Fig.~\ref{fig.flow chart}. For the sake of simplicity, we have made some adjustments to the process as compared to the methodology outlined in \cite{Lin}.

\begin{description}
    \item[Step 1:] We initialize the group of particles and configure key parameters within the code, such as the number of particles $N$, the dimension of the parameter space $D$, and the number of iterations $\mathcal{I}$.
    

    \item[Step 2:] We calculate the fitness for each particle, denoted as $f(\Vec{X}_i)$, and compare with their corresponding local best value, $Pbest_i$. If $f(\Vec{X}_i) < Pbest_i$, the local best position $\Vec{P}_i$ is updated to match the position of that particle, i.e., $\Vec{P}_i=\Vec{X}_i$.
    

    \item[Step 3:] We compare the local best value of all particles with the global best value, denoted by $Gbest$. If $Pbest_i < Gbest$, the global best position is updated: $\vec{G}=\vec{P}_i$.
    

    \item[Step 4:] Initiate the chaotic optimization, as detailed in Sect.~\ref{sec:chaos}, and compute the fitness of each chaotic particle, denoted as $f(\Vec{X}^{chaos}_i)$. If the obtained chaotic solution is superior, we update the local best positions $\Vec{P}_i$ similar to Step 3, otherwise we update the local best positions using the probability from Eq.~(\ref{eq.4.15}).

    \item[Step 5:] We calculate the mean best positions using Eq.~(\ref{eq.4.9}). The contraction-expansion factor $\gamma$ is gradually reduced in a linear manner as the iteration count increases. Then, all particle positions are updated using Eq.~(\ref{eq.4.11}).

    \item[Step 6:] We verify if the number of iterations has reached the maximum value. If the maximum value has not been reached, the algorithm returns to Step.2; otherwise, the algorithm concludes and outputs the result of the algorithm, $\Vec{G}$ and $Gbest$.
\end{description}

\section{Fisher matrix}
\label{sec:Fisher}
The particle swarm optimization algorithm, we described in the previous section, can only provide the best fit for our model as probed by some cosmological data. However, we usually want to know what are the confidence regions in the parameter space according to the errors in our measurements. In order to accomplish that, we are going to use the Fisher matrix information around our best fit values.

The Fisher matrix formalism offers a convenient and effective approach for examining the cosmological constraints derived from diverse data sets owing to its inherent simplicity. Given any $\chi^2$ function, we can expand it about its minimum at the parameter's values $\vec{X_0}$. Since the linear term vanishes at the minimum, the leading contribution for the deviation around the minimum is given by the quadratic term. The corresponding coefficients yields the Fisher matrix \citep{Coe:2009xf}
%
%
\begin{equation}
\label{eq.5.2}
\mathbf{F}=\frac{1}{2}\begin{pmatrix}
    \frac{\partial ^2}{\partial x_1\partial x_1}&   \frac{\partial ^2}{\partial x_1 \partial x_2}&  \cdots&  \frac{\partial ^2}{\partial x_1\partial x_n}\\
    \frac{\partial ^2}{\partial x_2\partial x_1}&   \frac{\partial ^2}{\partial x_2 \partial x_2}&  \cdots&  \frac{\partial ^2}{\partial x_2\partial x_n}\\
    \vdots& \vdots& \ddots& \vdots\\
    \frac{\partial ^2}{\partial x_n \partial x_1}&  \frac{\partial ^2}{\partial x_n \partial x_2}&  \cdots& \frac{\partial ^2}{\partial x_n\partial x_n}
\end{pmatrix}\chi ^2 \,,
\end{equation}
assuming a total of $n$ parameters in our model. 

The derivatives can be approximated using a discretization approach as follows:
\begin{equation}
\label{eq.5.3}
 \frac{\partial ^2\chi ^2}{\partial x_i^2}\approx\frac{\chi ^2(x_{i0}+\Delta x_i,\vec{X_0})-2\chi ^2(x_{i0},\vec{X_0})+\chi ^2(x_{i0}-\Delta x_i,\vec{X_0})}{\left(\Delta x_i\right)^2} \,,
\end{equation}
\begin{equation}
\label{eq.5.4}
 \frac{\partial \chi ^2}{\partial x_i}\approx\frac{\chi ^2(x_{i0}+\Delta x_i,\vec{X_0})-\chi ^2(x_{i0}-\Delta x_i,\vec{X_0})}{\left(\Delta x_i\right)^2} \,,
\end{equation}
\begin{equation}
\label{eq.5.5}
 \frac{\partial ^2\chi ^2}{\partial x_i\partial x_j}=\frac{\Delta \frac{\partial \chi ^2}{\partial x_i}}{\Delta x_j} \,,
\end{equation}
where $\vec{X_0}$ is a vector containing the best fit parameters other than $x_{i0}$ in $\chi^2$. The covariance matrix is obtained from the inverse of the Fisher matrix
\begin{equation}
\label{eq.5.6}
\mathbf{C}=\mathbf{F}^{-1}=
\begin{pmatrix}
    \sigma_{x_i}^2 & \sigma_{x_1x_2} & \cdots & \sigma_{x_1x_n} \\
    \sigma _{x_2x_1} & \sigma_{x_2}^2 & \cdots & \sigma_{x_2x_n} \\
    \vdots& \vdots& \ddots& \vdots \\
    \sigma_{x_nx_1} & \sigma_{x_nx_2} & \cdots & \sigma_{x_n}^2
\end{pmatrix} \,.
\end{equation}
And the computation of the confidence ellipses involving two parameters, such as $x_i$ and $x_j$, can be obtained as
%
%
\begin{equation}
\label{eq.5.7}
a^2=\frac{\sigma _{x_i}^2+\sigma _{x_j}^2}{2}+\sqrt{\frac{\left(\sigma _{x_i}^2-\sigma _{x_j}^2 \right)^2}{4}+\sigma _{{x_i}{x_j}}^2} \,,
\end{equation}
\begin{equation}
\label{eq.5.8}
b^2=\frac{\sigma _{x_i}^2+\sigma _{x_j}^2}{2}-\sqrt{\frac{\left(\sigma _{x_i}^2-\sigma _{x_j}^2 \right)^2}{4}+\sigma _{{x_i}{x_j}}^2} \,,
\end{equation}
\begin{equation}
\label{eq.5.9}
\tan2\theta=\frac{2\sigma _{x_ix_j}^2}{\sigma _{x_i}^2-\sigma_{x_j}^2} \,,
\end{equation}
where $a$ and $b$ denote half of the major and minor axes of the ellipse, respectively. The parameter $\theta$ indicates the angle by which the major axis deviates from the horizontal axis. We then multiply the axis lengths $a$ and $b$ by a coefficient $\alpha$ depending on the confidence level we are interested in. For instance, $\alpha = \sqrt{\Delta \chi^2} \approx 1.52$ for $68.3\%$ C.L. ($1\sigma$) and $\alpha = \sqrt{\Delta \chi^2} \approx 2.48$ for $95.4\%$ C.L. ($2\sigma$).


\section{Simulation}
\label{sec:Simulation}
We would like first to test the feasibility of the CQPSO algorithm and the Fisher matrix formalism to constrain our IDE models. Therefore, we build a simulation for the ages of 200 old astrophysical objects with known cosmology, which we use to check our ability to recover the fiducial model within the expected confidence level.

We consider two fiducial cosmologies. One with an interaction proportional to the energy density of dark energy (Model I) given by $\omega = -0.8$, $\xi = -0.2$, and $\Omega_c = 0.3$, where the density parameter for any component is defined as $\Omega_i \equiv (8\pi G/3H_0^2)\rho_{i0}$; and a second model with interaction proportional to the energy density of dark matter (Model II) with parameter values $\omega = -1.5$, $\xi = 0.2$, and $\Omega_c = 0.3$. For each of these two fiducial cosmologies, we generate 200 uniformly distributed points within the redshift range $z = (0,4]$. At each specific redshift $z_i$, we compute the age of the universe using Eq.~(\ref{eq.3.3}). Subsequently, we introduce a delay factor $df$ which is subtracted from the age of the universe at $z_i$, thereby determining the theoretical age of each observed astronomical object (OAO). For simplicity, we fix the delay factor $df = 0.2$ for all objects. We also assign $H_0$ and $\Omega_b$ to the corresponding best fit values with $\rm{LBT + H_0}$ in \cite{daCosta:2014kua}. 
Ultimately, the age of the OAO is generated by assuming a Gaussian distribution as depicted by
\begin{equation}
\label{Gaussian distribution}
\mathcal{P}(\mathscr{t}^{obs}_{i})=\frac{1}{\sqrt{2\pi}\sigma_i}\exp{\left(-\frac{[\mathscr{t}^{obs}_{i} - (\mathscr{t}_U(z_i) - df_i)]^2}{2\sigma_i^2}\right)} \,,
\end{equation}
where $\sigma_i$ is the error in our measurement, which we model as linearly decreasing with the corresponding redshift $z_i$ as follows:
\begin{equation}
\label{sigma-redshift}
\sigma_i = \sigma_0 + (0.01 - \sigma_0) \times \frac{z_i}{4} \,.
\end{equation}
We explore two scenarios with $\sigma_0=0.05 \, \rm{Gyr}$ and $\sigma_0=0.5 \, \rm{Gyr}$, respectively.

\begin{figure*}
\subfloat[]{
\includegraphics[width=0.47\textwidth]{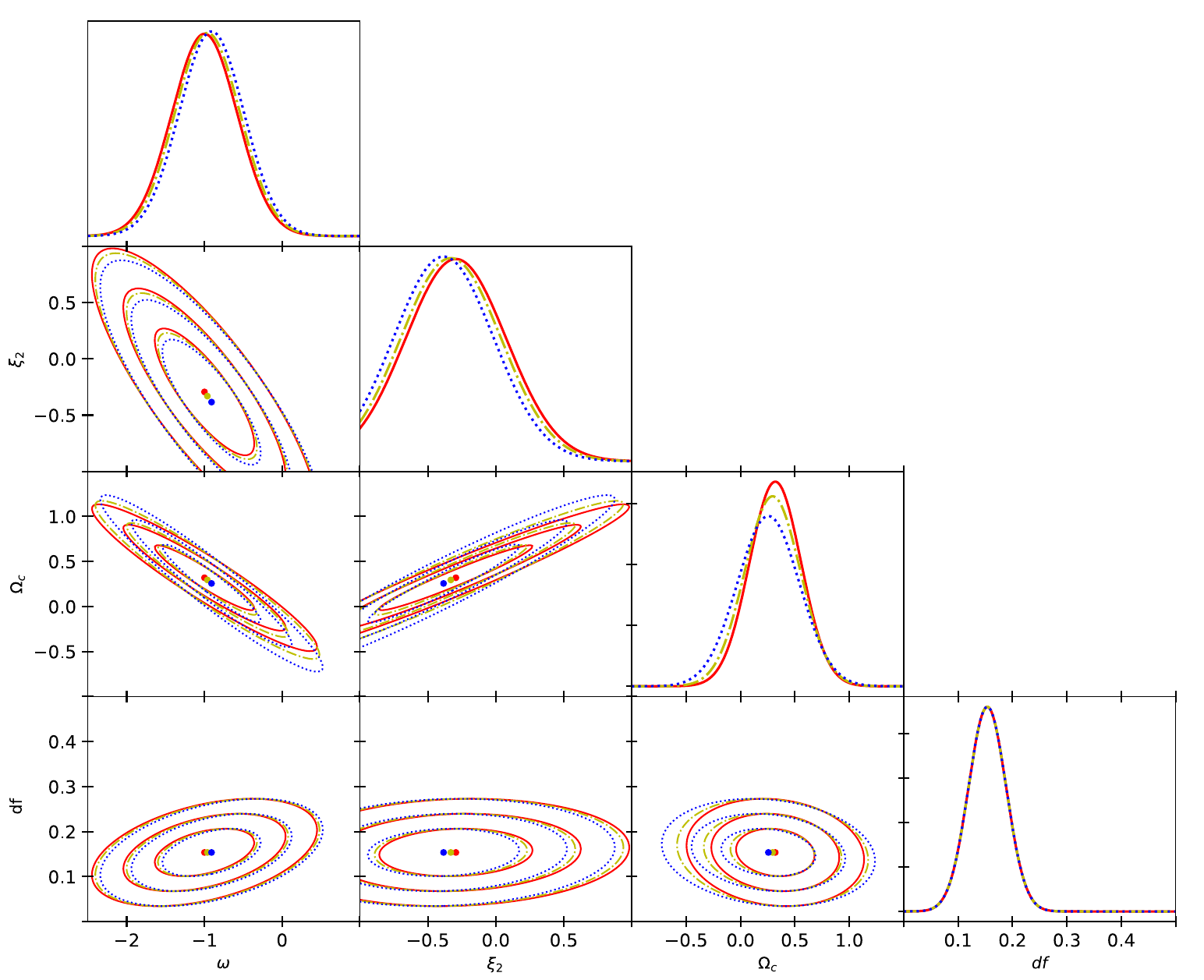}
}
\subfloat[]{
\includegraphics[width=0.47\textwidth]{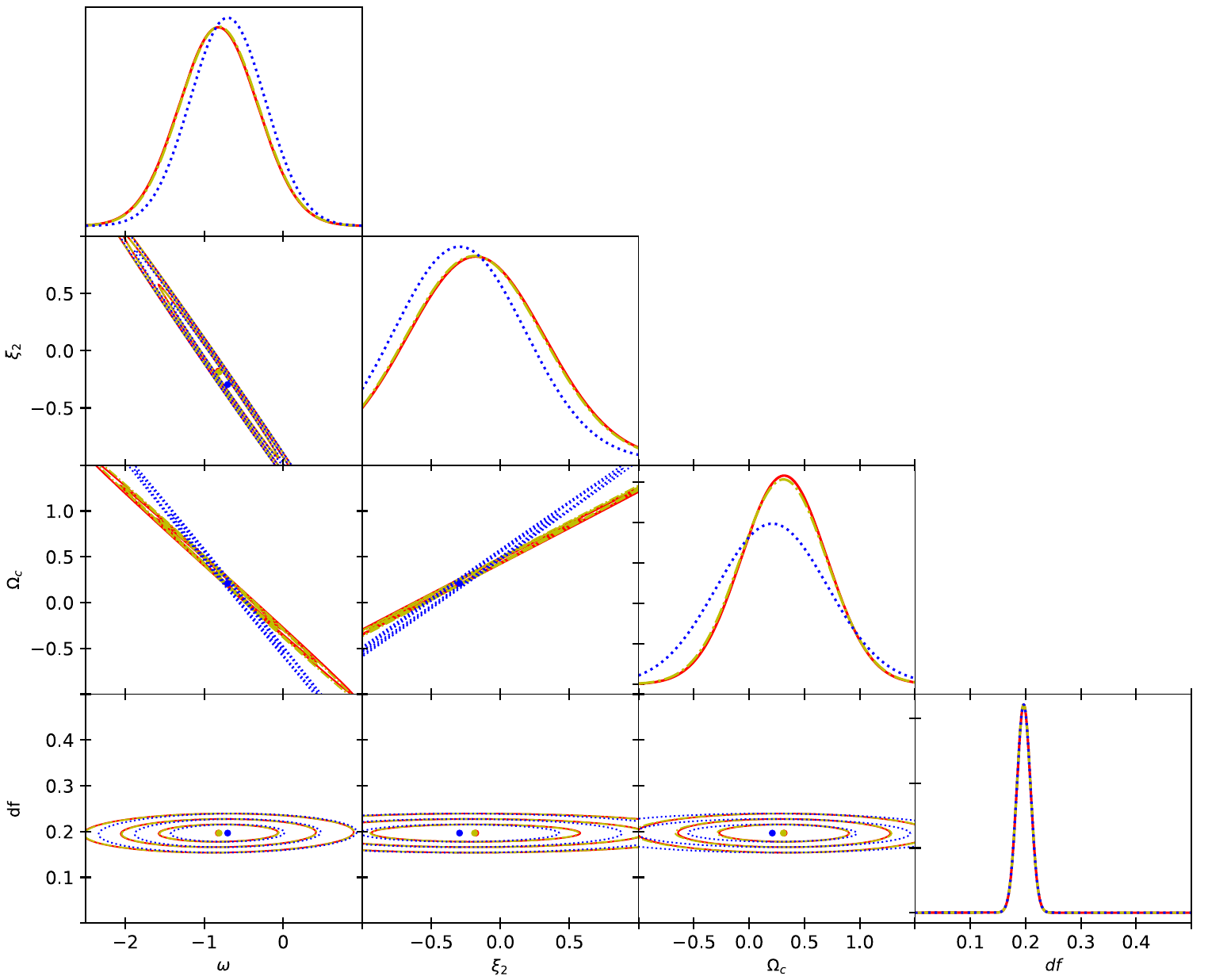}
}
\caption{1D and 2D marginalized posterior distributions with the
 $1 \sim 3 \sigma$ contours for simulations of Model~I.1 with (a) $\sigma_0=0.5$ and (b) $\sigma_0=0.05$.}
\label{fig.simulationI.1}
\end{figure*}

\begin{figure*}
\subfloat[]{
\includegraphics[width=0.47\textwidth]{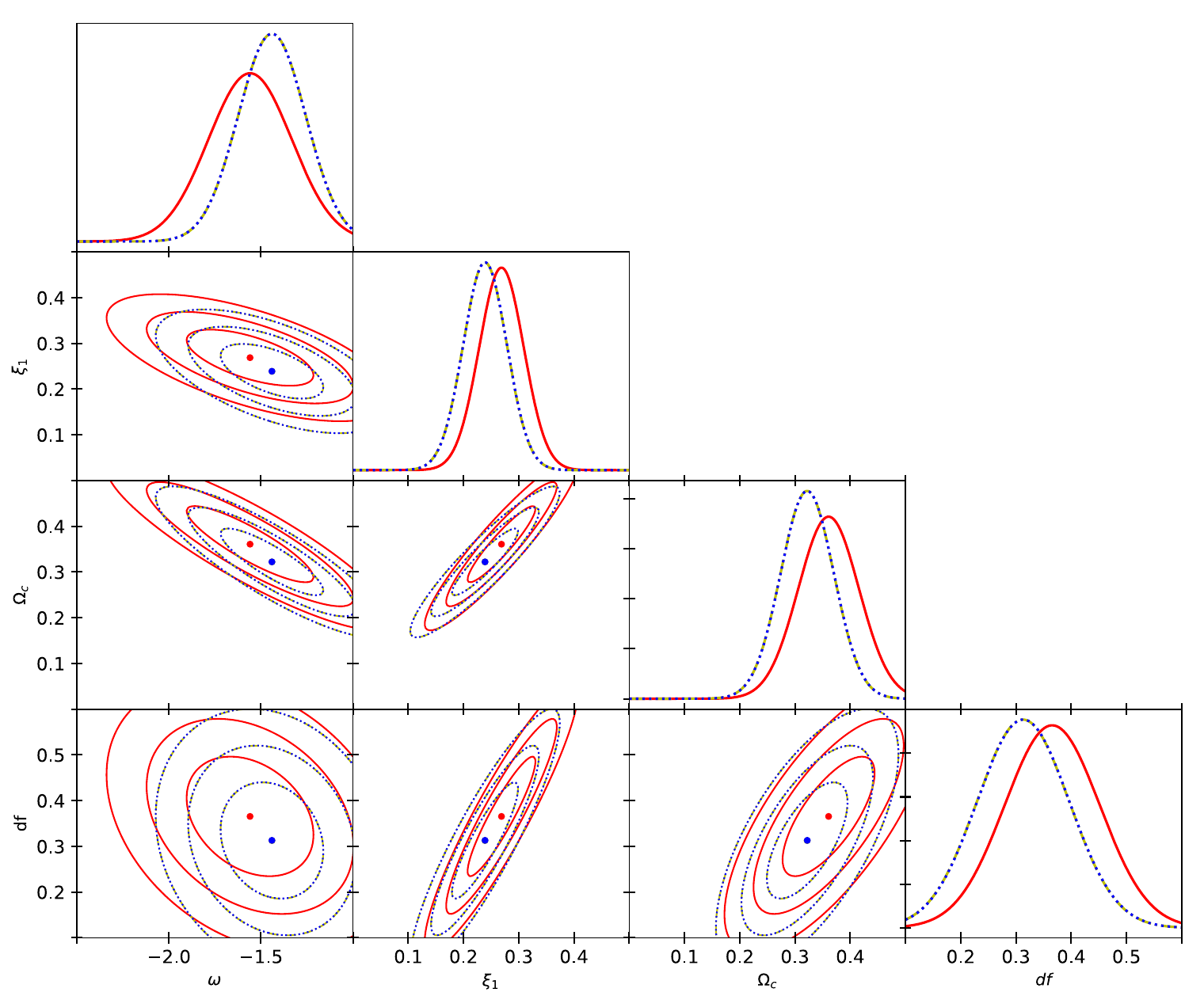}
}
\subfloat[]{
\includegraphics[width=0.47\textwidth]{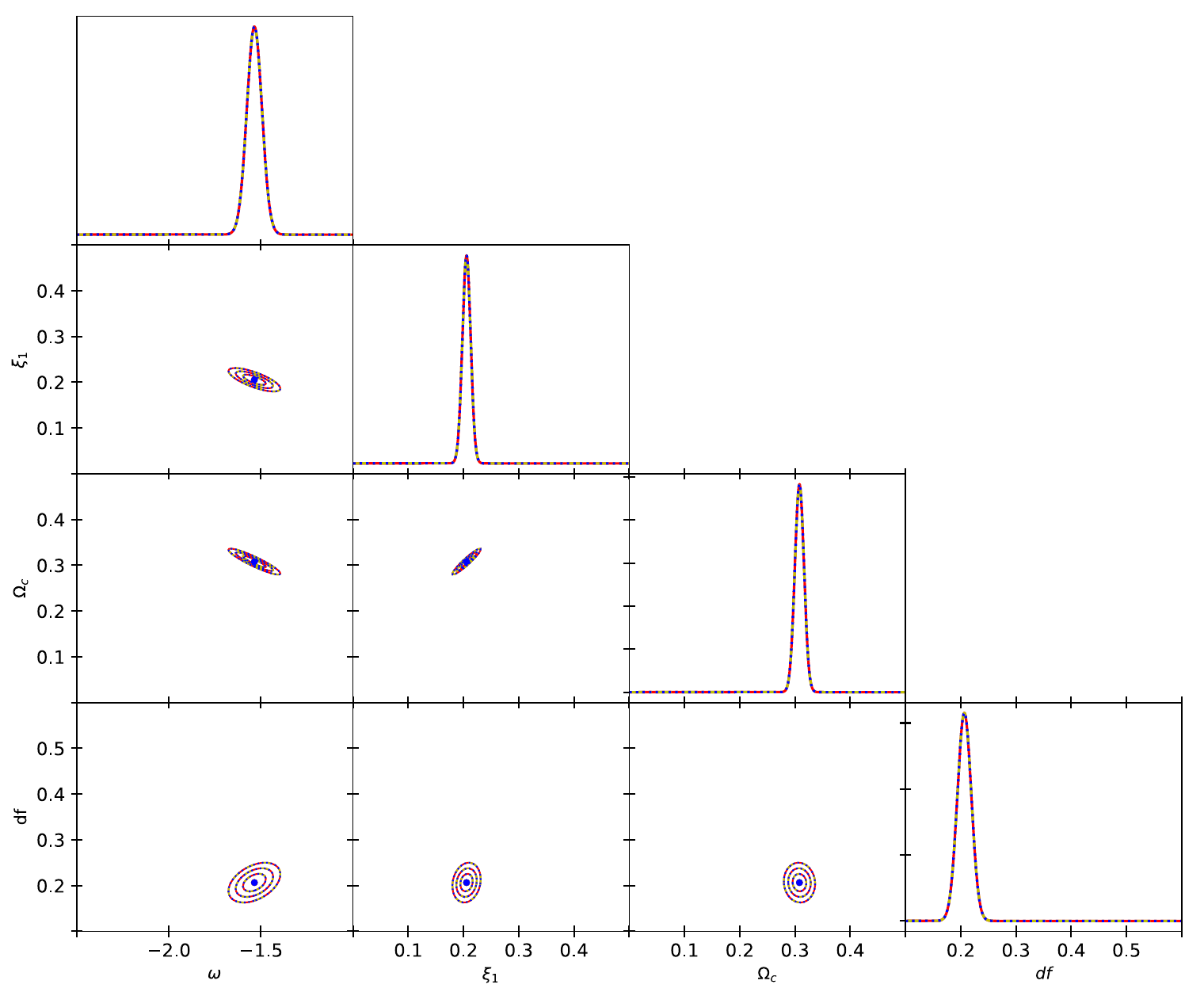}
}
\caption{1D and 2D marginalized posterior distributions with the
 $1 \sim 3 \sigma$ contours for simulations of Model~II with (a) $\sigma_0=0.5$ and (b) $\sigma_0=0.05$.}
\label{fig.simulationII}
\end{figure*}

For each of these simulations, we run the CQPSO algorithm and calculate the Fisher matrix to obtain the best fit and confidence levels. We set our prior as in Table~\ref{tab.prior}\footnote{Although the CQPSO method could also run with the more general case where all parameters are free, we observed a large degeneracy between our parameters, such that different runs of the CQPSO method could obtain different results with similar $\chi^2$. We therefore break some of these degeneracies by fixing $\Omega_bh^2$ and $H_0$ to the best bit values in \cite{daCosta:2014kua} for the same cosmological models using lookback time plus $H_0$ data. $\Omega_bh^2$ can actually be well constrained with CMB data and we also have strong constraints for $H_0$ from CMB or other low redshift data. Therefore, fixing those parameters can be seen as adding some additional data to break some degeneracies.}, which is separated into four models according to stability conditions as discussed in \cite{He:2008si}. Due to the inherent randomness of the CQPSO algorithm, we run the code three times to validate its performance. The outcomes of these runs are documented in Table~\ref{tab.simulation}, while the marginalized posterior distributions can be visualized in Figure~\ref{fig.simulationI.1} (for Model I.1) and Figure~\ref{fig.simulationII} (for Model II).
\begin{table}
    \centering
    \caption{Priors for the cosmological parameters considered in the analysis.}
    \begin{tabular}{|c|c|c|c|c|}
    \hline
        Parameters & \multicolumn{4}{c|}{Prior} \\ \hline
        $\Omega_{c0}$ & \multicolumn{4}{c|}{[0,1]} \\ \hline
        $df$ & \multicolumn{4}{c|}{[0,5]} \\ \hline
        ~ & Model I.1 & Model I.2 & Model II & Model III \\ \hline
        $\omega$ & [-1,-0.1] & [-2.5,-1] & [-2.5,-1] & [-2.5,-1] \\ \hline
        $\xi$ & [-0.4,0] & [0,0.4] & [0,0.4] & [0,0.4] \\ \hline
    \end{tabular}
    \label{tab.prior}
\end{table}




\begin{table*}
    \caption{Best fits and 68\% confidence level for our simulated data assuming two different fiducial cosmologies. For each model, we consider two cases with $\sigma_0 = 0.5$ and $\sigma_0 = 0.05$. And we run the CQPSO algorithm 3 times to check consistency. }
    \label{tab.simulation}
    \centering
    \begin{tabular}{cccccccccc}
        \bottomrule 
        \multicolumn{2}{c}{\centering \multirow{2}{*}{\centering \textbf{Model and Error}}} & \multicolumn{5}{c}{\textbf{Parameters}} \\ \cline{3-7}
        ~ & ~ & \textbf{$\omega$} & \textbf{$\xi$} & \textbf{$\Omega_c$} & \textbf{$df$} & \textbf{$\chi^2$} \\ \bottomrule
        ~ & Test 1 & $-0.9986 \pm 0.4211$ & $-0.2932 \pm 0.3706$ & $0.3196 \pm 0.2371$ & $0.1537 \pm 0.03466$ & $169.7$  \\ \cline{2-7} 
        \textbf{Model I.1 $\sigma_0$=0.5} & Test 2 & $-0.9613 \pm 0.4194$ & $-0.3306 \pm 0.3690$  & $0.2950 \pm 0.2553$  & $0.1537 \pm 0.03466$ & $169.7$ \\ \cline{2-7} 
        ~ & Test 3 & $-0.9079 \pm 0.4163$ & $-0.3840 \pm 0.3662$ & $0.2562 \pm 0.2849$ & $ 0.1537 \pm 0.03465$ & $169.7$  \\ \bottomrule
        ~ &  Test 1 & $-0.8202 \pm 0.4984$ & $-0.1779 \pm 0.4972$ & $0.3161 \pm 0.3871$ & $0.1965 \pm 0.0123$ & $117.6$ \\ \cline{2-7}
        \textbf{Model I.1 $\sigma_0$=0.05} & Test 2 & $-0.8116 \pm 0.4970$ & $-0.1866 \pm 0.4958$ & $0.3093 \pm 0.3943$ & $0.1965 \pm 0.0123$ & $117.6$ \\ \cline{2-7} 
        ~ & Test 3 & $-0.7028 \pm 0.4754$ & $-0.2954 \pm 0.4741$ & $0.2095 \pm 0.5029$ & $0.1965 \pm 0.01239$ & $117.6$  \\ \bottomrule
        ~ & Test 1 & $-1.559 \pm 0.2265$ & $0.2687 \pm 0.04037$ & $0.3608 \pm 0.05479$ & $0.3654 \pm 0.08602$ & $165.1$  \\ \cline{2-7} 
        \textbf{Model II $\sigma_0$=0.5} & Test 2 & $-1.440 \pm 0.1833$ & $0.2388 \pm 0.03933$ & $0.3219 \pm 0.04800$ & $0.3126 \pm 0.08365$ & $164.6$ \\ \cline{2-7} 
        ~ & Test 3 & $-1.440 \pm 0.1834$ & $0.2389 \pm 0.03933$ & $0.3221 \pm 0.04801$ & $0.3130 \pm 0.08364$ & $164.6$  \\ \bottomrule
        ~ & Test 1 & $-1.535 \pm 0.04097$ & $0.2055 \pm  0.007448$ & $0.3080 \pm 0.008246$ & $0.2065 \pm 0.01265$ & $110.7$  \\ \cline{2-7} 
        \textbf{Model II $\sigma_0$=0.05} & Test 2 & $-1.535 \pm 0.04097$ & $0.2055 \pm 0.007448$ & $0.3080 \pm 0.008246$ & $0.2065 \pm 0.01265$ & $110.7$ \\ \cline{2-7} 
        ~ & Test 3 & $-1.535 \pm 0.04097$ & $0.2055 \pm 0.007448$ & $0.3080 \pm 0.008246$ & $0.2065 \pm 0.01265$ & $110.7$  \\ \bottomrule
    \end{tabular}
\end{table*}

Analyzing the results presented in Table~\ref{tab.simulation}, it becomes evident that irrespective of the value chosen for $\sigma_0$, the best fits for the 3 runs generated by the CQPSO algorithm agree with each other within $1\sigma$ confidence level. In general, these values also agree with the fiducial cosmology we chose for our simulation, although the recovered delay factor $df$ is more than $1\sigma$ away when the error $\sigma_0 = 0.5 \, \rm{Gyr}$ in both Model I.1 and Model II. Comparing the results for Model I.1 and Model II, we observe a larger variation between the results for Model I.1, which we attribute to the fact that this model is proportional to the energy density of dark energy and, as such, is less susceptible to variations in the interaction. This is related to the sub-dominance of dark energy over a large fraction of the Universe history, as it was already discussed in previous works (e.g. \cite{Costa:2013sva,Costa:2016tpb}. It is also curious that the standard deviation for the parameters $\omega$, $\xi$, and $\Omega_c$ increase as $\sigma_0$ goes from $\sigma_0 = 0.5 \, \rm{Gyr}$ to $\sigma_0 = 0.05 \, \rm{Gyr}$ in Model~I.1. However, the constraint for the delay factor $df$ improves, and we have checked that the figure of merit (FoM) defined as the volume of the ellipsoid in the N-dimensional parameter space ($\rm{FoM} \equiv V \propto \rm{det} (F)^{-1/2}$), actually decreases. On the other hand, Model II possesses stronger constraints and its results are more consistent over the several tests as a consequence of its interaction depending on the energy density of dark matter.

Therefore, we have checked that using the CQPSO algorithm we are able to recover the fiducial cosmology and that our results are inside one standard deviation, which we have estimated from a Fisher matrix formalism. We can also see that the method works better depending on the character of our data, i.e. how precise is the data and how strongly it can break degeneracies among our different cosmological parameters, and also on the sensibility of our model about variations on its parameters.


\section{Real data}
\label{sec:Real}
We pass now to analyze our interacting dark energy models in face of real data from the ages of the OAO. First, we are going to use a set of 38 galaxies and cluster's ages and compare our results with some previous analysis using MCMC. This will also serve to test our method of analysis. Second, we increase this data set by employing new estimations for the ages of OAO, which amounts a total of 114 OAO objects.

\subsection{Analysis on 38 OAO ages}
We delve deeper into evaluating the feasibility of CQPSO using a data source of 32 passively evolving galaxies in the redshift interval $0.117 \leq z \leq 1.845$ with an uncertainty in the age measurements of $12\%$ at one standard deviation as presented in \cite{Samushia:2009px}. In addition, we use the ages of 6 galaxy clusters between $0.10 \leq z \leq 1.27$ and an uncertainty of $1 \, \rm{Gyr}$ \citep{Capozziello:2004jy}. We then explore the parameter space using the CQPSO algorithm for the likelihood given by Eq.~(\ref{Eq.3.6}). We consider Model I.2 and Model II with priors described in Table~\ref{tab.prior}. Because of the degeneracy between our parameters and the weakness of our data to properly constrain altogether, we fix $\Omega_bh^2 = 0.1355$ and $H_0 = 74.12$ in Model I.2, and $\Omega_bh^2 = 0.009584$ and $H_0 = 73.25$ in Model II. These are the best-fit values we obtained previously using a MCMC analysis \citep{daCosta:2014kua}. Although fixing these values restrict our analysis, they could be considered as constraints from other data set. Our main motivation here is to re-obtain the results in \cite{daCosta:2014kua} using the CQPSO method.

\begin{figure*}
\subfloat[]{
\includegraphics[width=0.47\textwidth]{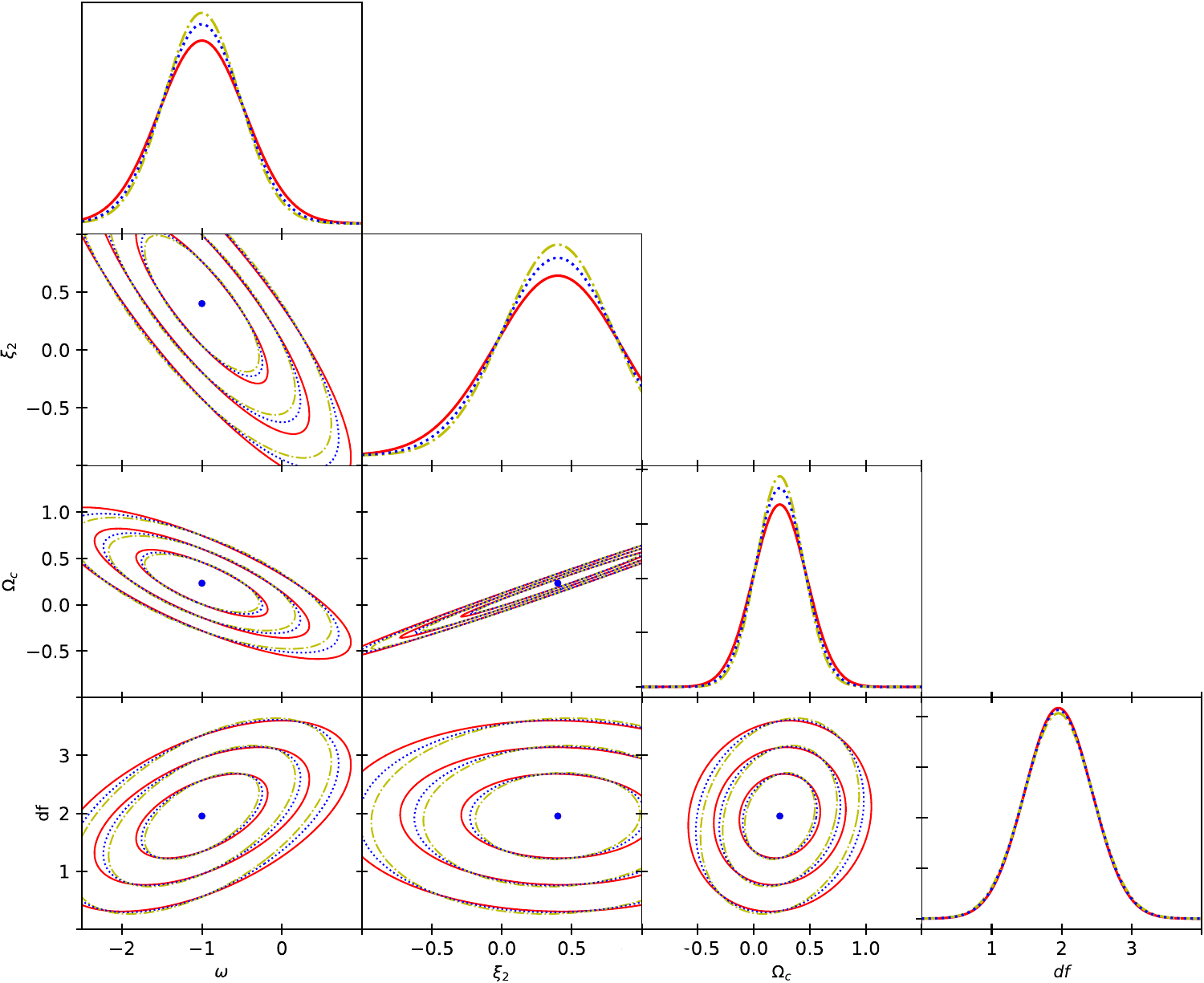}
}
\subfloat[]{
\includegraphics[width=0.47\textwidth]{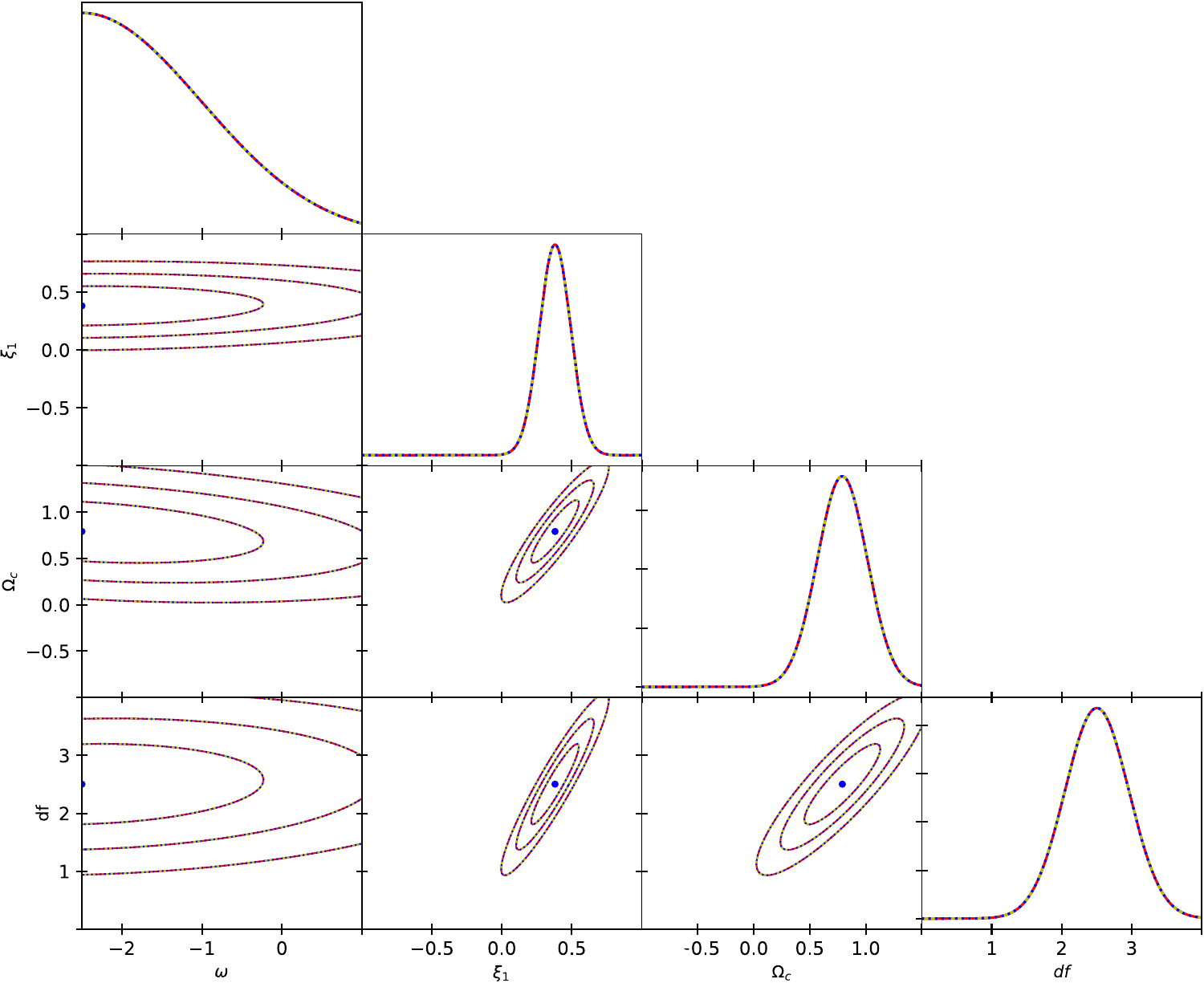}
}
\caption{1D and 2D marginalized posterior distributions with the
 $1 \sim 3 \sigma$ contours for the real data with a total of 38 galaxies and clusters in (a) Model~I.2 and (b) Model~II.}
\label{real38}
\end{figure*}

We run the CQPSO algorithm several times as we did in Sect.~\ref{sec:Simulation}. All results for the best fit were in very good agreement with each other by at least 4 significant figures. The average result and the corresponding standard deviation, which we obtain from the Fisher matrix, is presented in Table~\ref{tab.38data}. Figure~\ref{real38} show the 1D and 2D marginalized posterior distributions for 3 different runs in each model.
\begin{table*}
    \caption{Best fits and 68\% confidence level for the real data with a total of 38 galaxies and clusters. We show the average results for 2 interacting models. }
    \label{tab.38data}
    \centering
    \begin{tabular}{cccccccccc}
        \bottomrule 
        \multicolumn{2}{c}{\centering \multirow{2}{*}{\textbf{Models}}} & \multicolumn{5}{c}{\textbf{Parameter}} \\ \cline{3-7}
        ~ & ~ & \textbf{$\omega$} & \textbf{$\xi$} & \textbf{$\Omega_c$} & \textbf{$df$} & \textbf{$\chi^2$} \\ \bottomrule
        \textbf{Model I.2} &  & $-1.0 \pm 0.5$ & $0.4 \pm 0.4$ & $0.23 \pm 0.22$ & $1.95 \pm 0.48$ & $27.79$ \\ 
        \bottomrule 
        \textbf{Model II} &  & $-2.5 \pm 1.5$ & $0.38 \pm 0.11$  & $0.79 \pm 0.22$  & $2.50 \pm 0.46$ & $23.15$ \\
        \bottomrule
    \end{tabular}
\end{table*}

Let us now compare our results with those in Table~4.8 and Table~4.9 of \cite{daCosta:2014kua}. There the best fit for Model I.2 is given by: $\Omega_bh^2 = 0.1355$, $\Omega_ch^2 = 0.1255$, $H_0 = 74.12$, $\omega = -1.015$, $\xi = 0.398$, $df = 1.937$, and $\chi^2 = 28.06$. As we can see by comparing these results with those in Table~\ref{tab.38data}, they are in very good agreement and the CQPSO method with 4 free parameters was able to recover the results obtained with the MCMC. On the other hand, our results for their standard deviations are $1.4$ up to $4$ times larger than the MCMC result. Their best-fit results for Model II are: $\Omega_bh^2 = 0.009584$, $\Omega_ch^2 = 0.386$, $H_0 = 73.25$, $\omega = -1.558$, $\xi = 0.3955$, $df = 2.554$, and $\chi^2 = 23.18$. Comparing with our results in Table~\ref{tab.38data}, we observe the $\chi^2$ is very similar and all parameters fall inside our $1\sigma$ confidence level, especially $\xi$ and $df$ which are very close to their results. The equation of state of dark energy seems quite offset, although still inside $1\sigma$, however Figure~4.13 in \cite{daCosta:2014kua} shows that the posterior from the MCMC for $\omega$ is almost flat, which we could not properly obtain from our fisher matrix analysis. Our standard deviations were 1.4 up to 2 times larger than the MCMC result.

\subsection{Analysis on 114 OAO ages}
Our simulation in Sect.~\ref{sec:Simulation} and also the comparison with previous results using MCMC for an old data set of 38 OAO made us confident that our method using the CQPSO can be used to find the best fits. In addition, although our standard deviations were larger than the MCMC result, they can be seen as more conservative. Therefore, we now analyse a new data set for the ages of 114 OAO. This data set encompasses galaxies and quasars (QSOs) detected up to approximately $z = 8$, the primary source of galaxy data is from the Cosmic Assembly Near-infrared Deep Extragalactic Legacy Survey (CANDELS) observational initiative \citep{Grogin:2011ua,2019ApJS..243...22B,2015ApJ...801...97S,2017ApJS..228....7N}. The data set encompasses the age-redshift of 61 galaxies and 53 quasars, following the selection in \cite{Vagnozzi:2021tjv}. We set the age of the Universe with the best fit values and uncertainties as reported in \cite{daCosta:2014kua} from Planck satellite for each respective model.
\begin{table*}
    \caption{Best fits and 68\% confidence level for the real data with 114 OAO. We show the average results for 4 interacting dark energy models and 3 different configurations for the data set.}
    \label{tab.114data}
    \centering
    \begin{tabular}{cccccccccc}
        \bottomrule 
        \multicolumn{2}{c}{\centering \multirow{2}{*}{\centering \textbf{Data and Model}}} & \multicolumn{5}{c}{\textbf{Parameter}} \\ \cline{3-7}
        ~ & ~ & \textbf{$\omega$} & \textbf{$\xi$} & \textbf{$\Omega_c$} & \textbf{$df$} & \textbf{$\chi^2$} \\ \bottomrule
        ~ &  Model I.1 & $ -1.0 \pm 0.42 $ & $ -0.37 \pm0.34 $ & $ 0  \pm 0.32  $ & $ 0.42 \pm 0.12 $ & $27.02$ \\ \cline{2-7}
        \textbf{All together} & Model I.2 & $ -1.4 \pm 0.51 $ & $ 0.13 \pm 0.39 $ & $ 0.093 \pm 0.187 $ & $ 0.42 \pm0.14 $ & $25.60$ \\ \cline{2-7} 
        ~ &  Model II & $ -2.5 \pm 1.34 $ & $ 0.155 \pm 0.093 $ & $ 0.45 \pm 0.13  $ & $ 0.73 \pm 0.38 $ & $15.09$ \\ \cline{2-7}
        ~ & Model III & $ -2.5 \pm 1.16 $ & $ 0.103 \pm0.042 $ & $ 0 \pm 0.74$ & $ 0.58 \pm 0.15 $ & $26.74$  \\ \bottomrule
        ~ &  Model I.1 & $ -1.0 \pm 0.31 $ & $ -0.38 \pm 0.30 $ & $ 0  \pm 0.23 $ & $ 0.16 \pm 0.17 $ & $6.595$ \\ \cline{ 2-7 }
        \textbf{Galaxy} & Model I.2  & $ -2.5 \pm 1.11 $ & $ 0.07 \pm 0.39 $ & $ 0.11 \pm 0.099 $ & $ 0 \pm 0.17 $ & $2.799$ \\ \cline{ 2-7 } 
        ~ &  Model II & $ -2.5 \pm 0.98 $ & $ 0 \pm 0.16 $ & $ 0.32 \pm 0.12 $ & $ 0.079 \pm 0.54 $ & $2.979$ \\ \cline{ 2-7 }
        ~ & Model III & $ -2.5 \pm 0.88 $ & $ 0.102 \pm 0.044 $ & $ 0 \pm 0.78 $ & $ 0.31 \pm 0.21 $ & $4.198$  \\ \bottomrule
        ~ &  Model I.1  & $ -1.0 \pm 0.31 $ & $ -0.37 \pm 0.29 $ & $ 0  \pm 0.23 $ & $ 0.44 \pm 0.11 $ & $8.498$ \\ \cline{ 2-7 }
        \textbf{Quasar} & Model I.2  & $ -2.5 \pm 1.06 $ & $ 9.0 \times 10^{-4} \pm 0.39 $ & $ 0.095 \pm 0.102 $ & $ 0.35 \pm 0.12 $ & $4.909$ \\ \cline{ 2-7 } 
        ~ &  Model II & $ -2.5 \pm 1.25 $ & $ 0 \pm 0.15 $ & $ 0.31 \pm 0.12 $ & $ 0.39 \pm 0.43 $ & $1.945$ \\ \cline{ 2-7 }
        ~ & Model III & $ -2.5 \pm 1.02 $ & $ 0.102 \pm 0.033 $ & $ 0 \pm 0.59 $ & $ 0.63 \pm 0.13 $ & $10.33$  \\ \bottomrule

    \end{tabular}
\end{table*}

\begin{figure*}
\subfloat[]{
\includegraphics[width=0.47\textwidth]{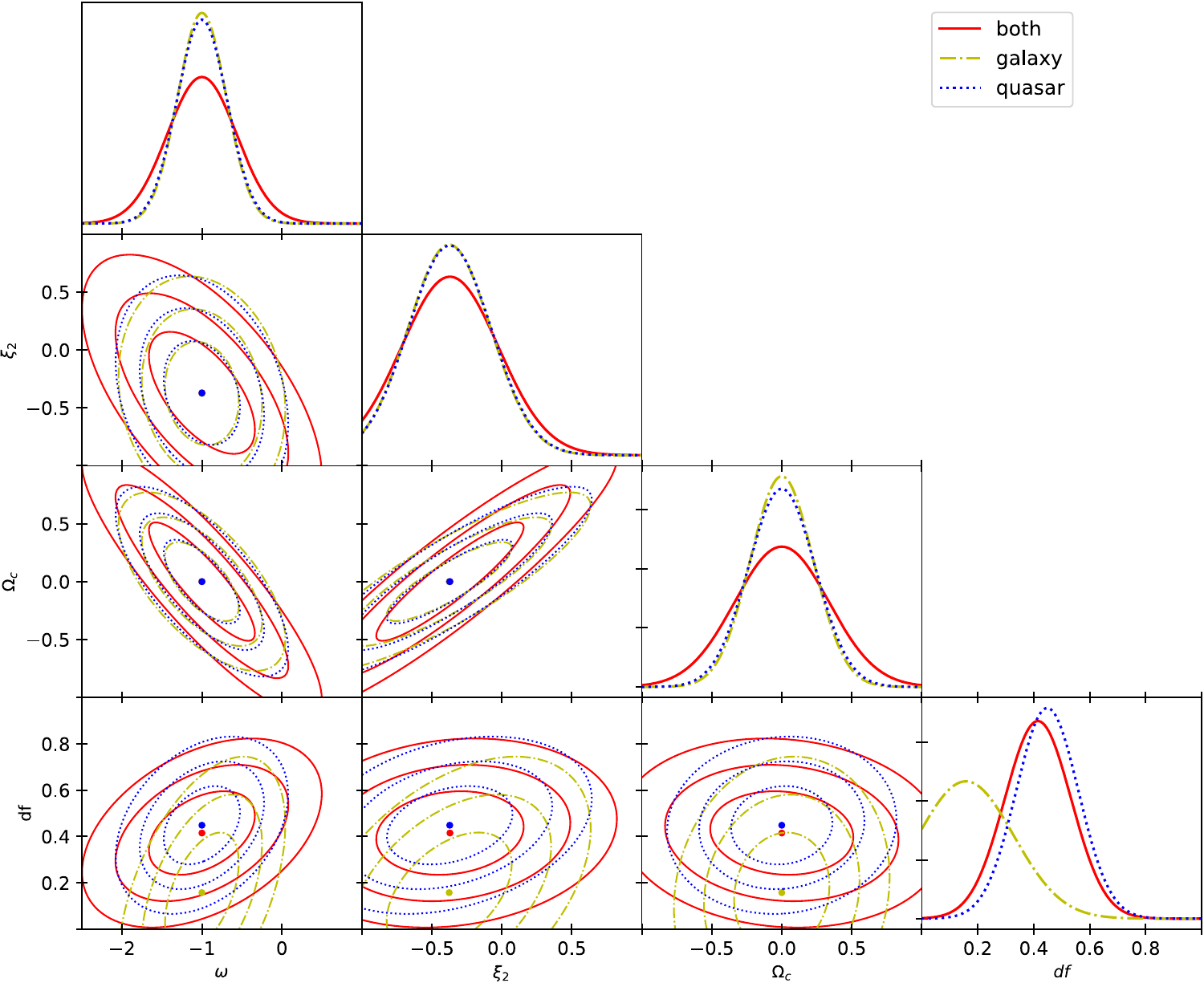}
}
\subfloat[]{
\includegraphics[width=0.47\textwidth]{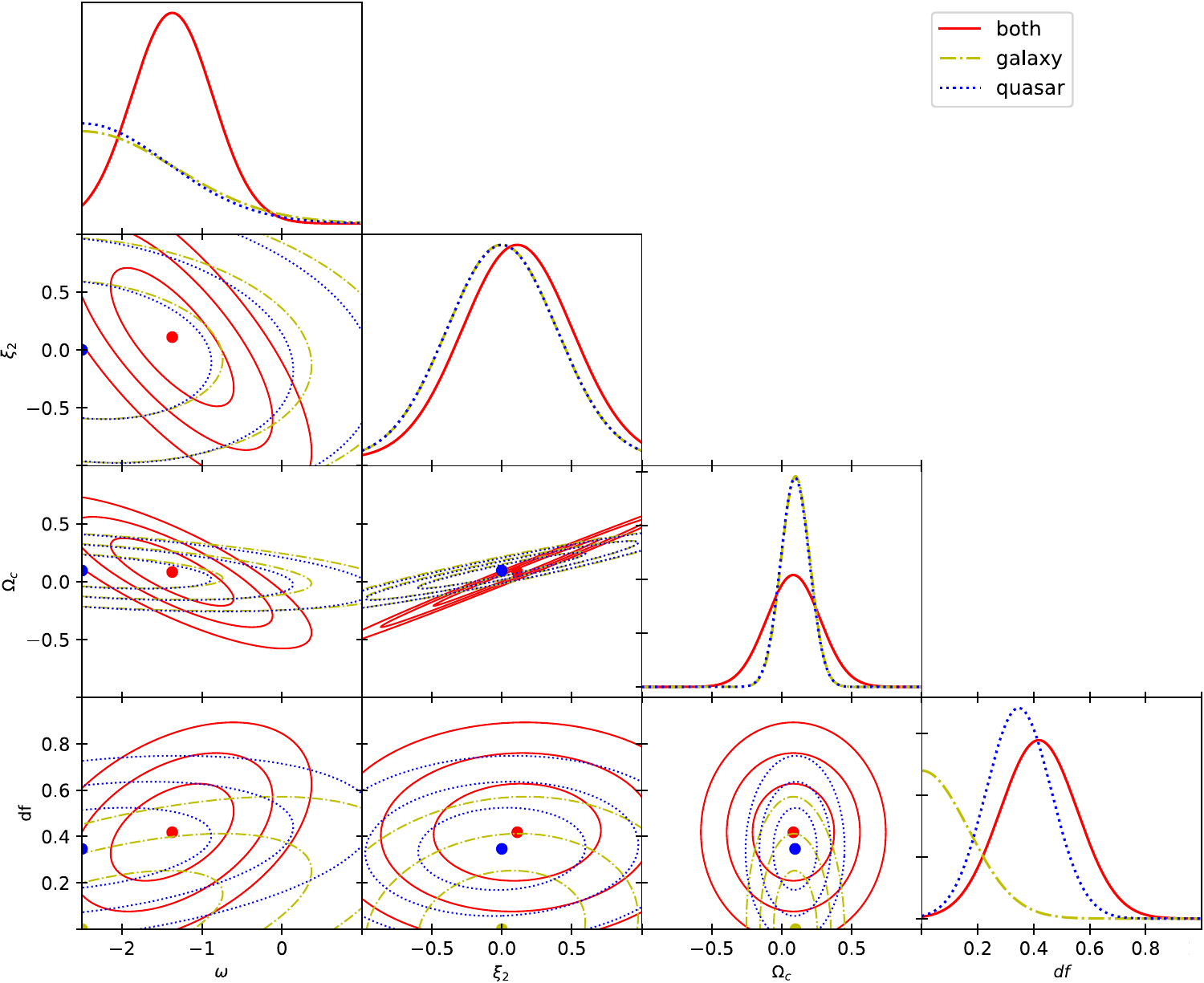}
}
\\
\subfloat[]{
\includegraphics[width=0.47\textwidth]{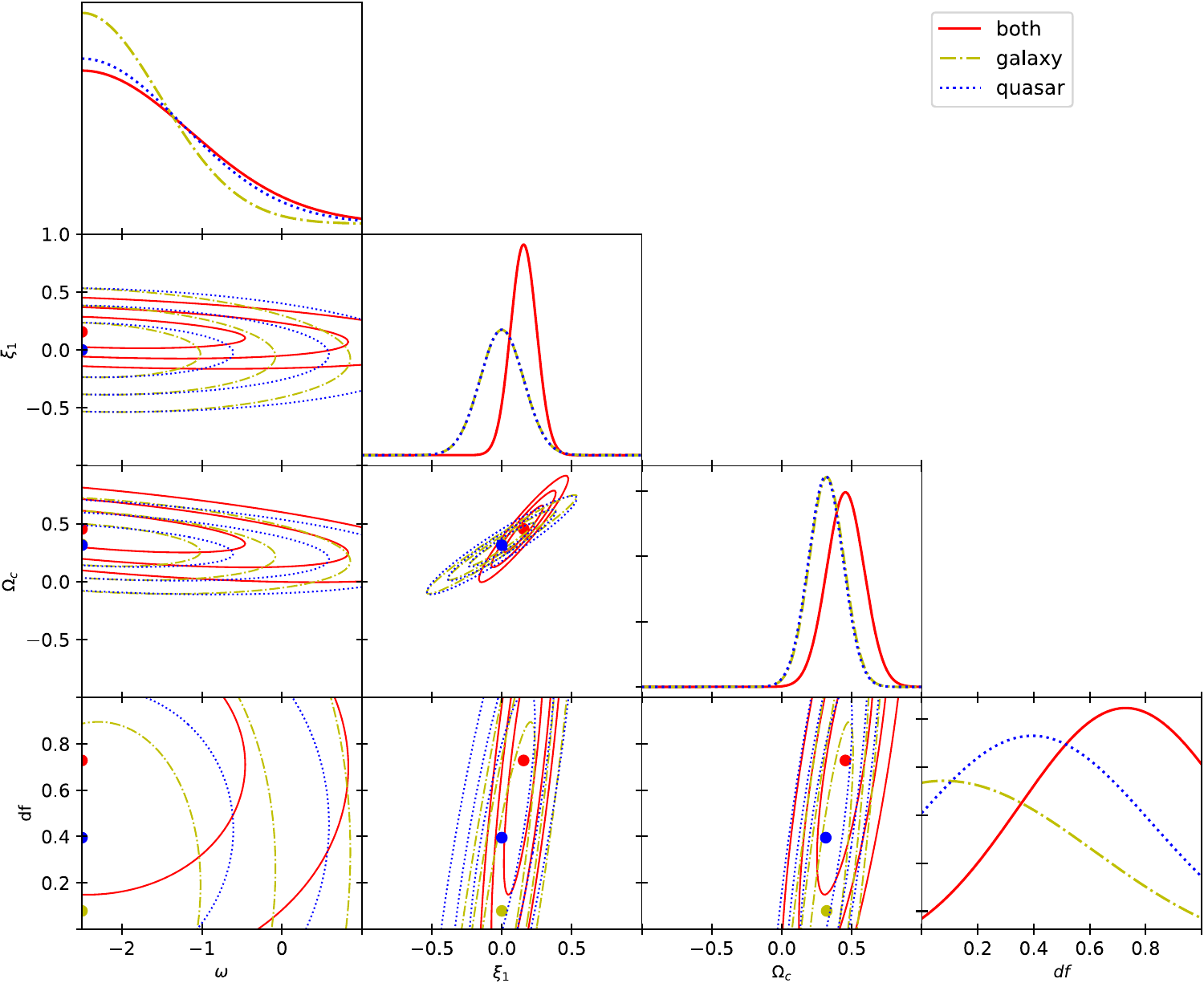}
}
\subfloat[]{
\includegraphics[width=0.47\textwidth]{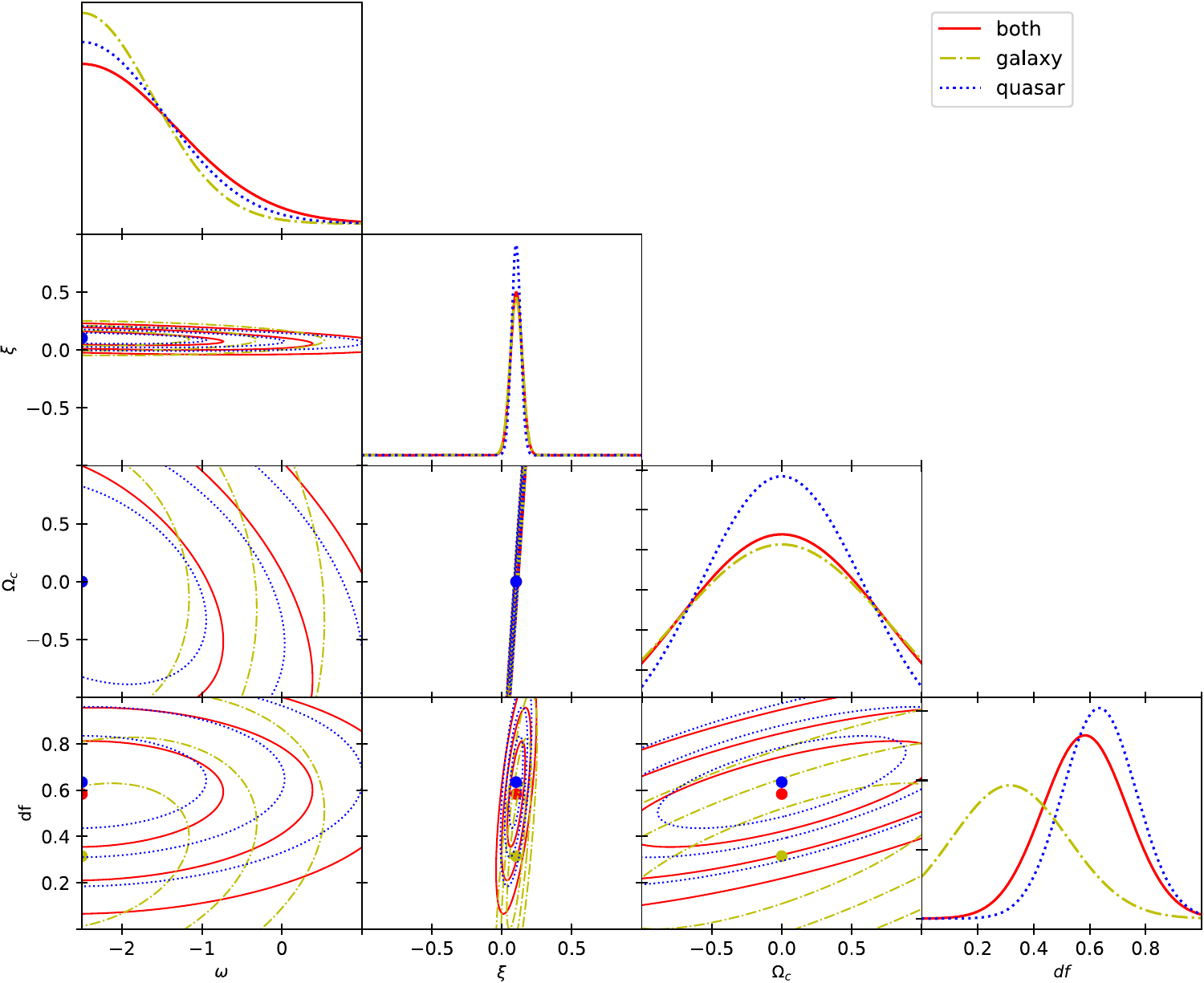}
}
\caption{1D and 2D marginalized posterior distributions with the $1 \sim 3 \sigma$ contours for the real data of 114 OAO ages in (a) Model~I.1, (b) Model~I.2, (c) Model~II, and (d) Model~III.}
\label{real114}
\end{figure*}

Now we consider all four interacting dark energy models with priors described in Table~\ref{tab.prior}. As before, because of the degeneracy between our parameters and the inability of our data to disentangle them appropriately, we have fixed $\Omega_bh^2$ and $H_0$ to the best fit values obtained in \cite{daCosta:2014kua}. We again ran the CQPSO algorithm several times to check consistency and observed a good agreement among them, well within $1\sigma$ confidence level. Table~\ref{tab.114data} presents the best fit and $68\%$ confidence level for the average of several runs in each case. We considered 3 different configurations for our data: 1) Combining all galaxies and quasars; 2) Considering only galaxies; 3) Considering only quasars. As we observed in \cite{Costa:2023cmu}, the ages of quasars are not consistent with the ages of galaxies in this data set, therefore, we analysed these 3 situations. Figure~\ref{real114} shows the 1D and 2D marginalized posterior distribution for each scenario.

Analysing the results for Model~I.1, we first see that, independent of the data set used, the equation of state of dark energy tend to $\omega = -1$. Second, the interaction parameter tend to large negative values, with a best fit $\xi \sim 0.37$. Here, it is important to note that although the error bars are large, they exclude a null interaction at $68\%$ confidence level. Third, the amount of dark matter tend to the lowest possible values. Finally, the best fit for the delay factor is $df = 0.16$ with galaxies and $df = 0.44$ with quasars, and they are not compatible at $68\%$ confidence level.

In Model~I.2, both galaxies and quasars reproduce similar results for the equation of state of dark energy, which tend to large negative values $\omega = -2.5$. The interaction parameter is consistent with $\xi = 0$, and $\Omega_c \sim 0.1$. However, once again their delay factors are not compatible, with a best fit of $df = 0$ for galaxies and $df = 0.35$ for quasars. Therefore, combining these two data sets move the delay factor to $df = 0.42$ and shift the equation of state to $\omega = 1.4$. The best fit for the interaction also becomes larger, but the error bar is still in agreement with a null interaction. $\Omega_c$ remains basically the same, with a larger error bar.

Model~II also has a best fit for the equation of state of dark energy tending to $\omega = -2.5$ in every scenario. The interaction parameter tend to $\xi = 0$ for galaxies and quasars, but it is shifted to $\xi = 0.155 \pm 0.093$ combining those two data sets. Once again, this difference arises from incompatibility between these two data sets, as can be seen in the best fit for the delay factor. $\Omega_c \sim 0.31$ using galaxies or quasars, and combining both data the best fit is $\Omega_c = 0.45$.

Finally, Model~III has best fit $\omega = -2.5$, $\xi = 0.10$ and excludes the null condition, $\Omega_c$ tends to zero, but with large error bars, and the delay factor $df = 0.31$ (galaxies), $df = 0.63$ (quasars), and $df = 0.58$ (all together).

It is important to note that the $\chi^2$ using these new data from galaxies and quasars are much smaller than the result with the previous data in Table~\ref{tab.38data}, even though they are more numerous. This is related to larger error bars in our measurement and stronger consistency among the new data, which leads to an over fitted sample. On the other hand, combining these two data from galaxies and quasars significantly increases the $\chi^2$, which also indicates they are not compatible.

\section{Conclusions}
\label{sec:Conclusions}
In this paper, we have obtained, for the first time, an analytical solution for the most general interacting dark energy model with a phenomenological energy transfer $\mathcal{E} = 3H(\xi_1\rho_c + \xi_2\rho_d)$. This greatly simplifies the calculation for the evolution of dark energy and dark matter under these models, which had only been obtained numerically. It also improves the computational time, which otherwise needs loops and interpolations to properly match the functions in the coupled system. This can be very helpful in later MCMC constraints.

We also use a new data set with the age-redshift relation for 114 OAO and constrain the cosmological parameters for 4 interacting dark energy scenarios. We show the interaction is favored in some cases, although the uncertainty is still large. Comparing with previous results using Planck and other low redshift measurements as in \cite{Costa:2013sva} and \cite{Costa:2016tpb}, we obtain compatible results, although their uncertainty is one order of magnitude stronger or even larger. Combining these OAO with other measurements could break some degeneracies and provide complementary information on the evolution of dark energy and dark matter.

We use a method inspired in artificial intelligence, known as {\it Chaos Quantum-Behaved Particle Swarm Optimization} (CQPSO), to find the best fit values for our cosmological parameters in those interacting models. We test this method with both a controlled simulation and also comparing the results obtained from an old data set with a previous MCMC analysis. We show the method was able to recover the expected best fits, although we needed to fix $\Omega_bh^2$ and $H_0$ to break some degeneracies in the parameter space, which our lookback time data was not able to break alone. This method cannot properly sample the parameter space, therefore it is not intended to substitute a more complete MCMC analysis. In fact, in order to obtain the uncertainties in our parameters, we resort to the Fisher matrix formalism. But it can be used for a faster analysis and could also be used in combination with a Fisher matrix to make forecasts, or even in combination with an MCMC analysis searching for the position of the best fit values.

\section*{Acknowledgements}
We would like to express our sincere thanks to Dr. Jun-Jie Wei for sharing the compilation of the age-redshift data from OAO with us. A.A.C also acknowledges financial support from the National Natural Science Foundation of China (grant 12175192).

\section*{Data Availability}
The data underlying this article will be shared on reasonable request to the corresponding author.



\bibliographystyle{mnras}

\bibliography{references} 








\bsp	
\label{lastpage}
\end{CJK}
\end{document}